\documentclass[pss]{wiley2sp} 
\usepackage{amsmath}
\usepackage{soul}
\usepackage{graphicx}
\usepackage{epstopdf}

\def\be{\begin{equation}}
\def\ee{\end{equation}}
\def\ba{\begin{eqnarray}}
\def\ea{\end{eqnarray}}

\begin{document}

\title{Constraints on the total coupling strength 
to bosons in the iron based superconductors.}

\titlerunning{Constraints on the total coupling strength ... }

\author{%
  Stefan-Ludwig\ Drechsler\textsuperscript{\Ast,\textsf{\bfseries 1}},
  Helge Rosner\textsuperscript{\textsf{\bfseries 2}},
  Vadim Grinenko\textsuperscript{\textsf{\bfseries 4}},
  Steve Johnston\textsuperscript{\textsf{\bfseries 1,3}}
}

\authorrunning{Stefan-Ludwig\ Drechsler {\it et al.}}

\mail{e-mail: 
  \textsf{s.l.drechsler@ifw-dresden.de}}

\institute{%
  \textsuperscript{1}\,Leibniz-Institute for Solid State and Materials Research, Institute for Solid State Theory,
  IFW-Dresden, D-01171 Dresden, Germany\\
 \textsuperscript{2}\,Max-Planck-Institute for Chemical Physics of Solids, Dresden, Germany\\
\textsuperscript{3}\, Department of Physics and Astronomy, University of Tennesee, Knoxville 37996, USA\\
 \textsuperscript{4}\, Technical University of Dresden, Institute for Solid State Physics, Dresden, Germany\\
  }

\received{XXXX, revised XXXX, accepted XXXX} 
\published{XXXX} 


\abstract{%
Despite the fact that the Fe-based superconductors (FeSCs) were discovered nearly ten years ago, the community is still devoting a tremendous effort towards elucidating their relevant microscopic pairing mechanism(s) and interactions.  At present, there is still no consistent interpretation of their normal state properties, where the strength of the electron-electron interaction and the role of correlation effects are under debate. Here, we examine several common materials and illustrate various problems and concepts that are generic for all FeSCs. Based on empirical observations and qualitative insight from density functional theory, we show that the superconducting and low-energy thermodynamic properties of the FeSCs can be described semi-quantitively within multiband Eliashberg theory.  We account for an important high-energy mass renormalization phenomenologically,
and in agreement with constraints provided by thermodynamic, optical, and 
angle-resolved photoemission data. When seen in this way, all FeSCs 
with $T_\mathrm{c} <$ 40~K studied so far are found to belong to an
 {\it intermediate} coupling regime.  This finding is in contrast to the 
 strong coupling scenarios proposed in the early period of the FeSC history. 
 We also discuss several related issues, including the role of band shifts 
 as measured by the positions of van Hove singularities, and the nature of 
 a recently suggested quantum critical point in the strongly hole-doped 
 systems AFe$_2$As$_2$ (A = K, Rb, Cs).  Using high-precision full relativistic 
 GGA-band structure calculations, we arrive at a somewhat milder mass 
 renormalization in comparison with previous studies. From the calculated mass 
 anisotropies of all Fermi surface sheets, 
only the $\varepsilon$-pocket near the corner of the BZ is compatible 
with the experimentally observed anisotropy of the upper critical field, 
pointing to its dominant role in the superconductivity of these three compounds.
}

\maketitle
\section{Introduction.}
The physics of the Fe-based superconductors (FeSCs), 
discovered first in the ``precursor" low-$T_c$ material LaFeOP in
2006 \cite{Kamihara2006} and followed by the F-doped pnictide
isomorphic sister compound in 2008 \cite{Kamihara2008}, is very rich and
somewhat distinctive in comparison to other families of conventional and
unconventional superconductors
\cite{Johnston2010,Paglione2010,Stewart,Hirschfeld2016,Mannella2014}. In
particular, it is still unclear whether a single pairing mechanism with some
modifications is at work in these materials (e.g.\ mediated by interband
spin-fluctuations), or if charge and orbital fluctuations or phonons also play
some role.  (The latter were initially excluded both theoretically by density
functional theory [DFT] approaches, which ignore magnetic, orbital, and
correlation effects \cite{Boeri2008}, and experimentally by femtosecond
time-resolved photoemission measurements \cite{Rettig2013}, both yielding
electron-phonon [el-ph] coupling constants $\lambda_{\rm \tiny el-ph} \leq
0.2$.) Related questions  concern how to classify the sometimes observed nodal
superconducting order parameter (SCOP) as either an accidental $s_{ \pm}$
\cite{Hardy2013,Hardy2016} symmetry or as a protected $d$-wave symmetry
\cite{Reid2012,Grinenko2013,Kim2014}.  
Or, is there a change in symmetry of the SCOP from a sign-reversing to a
sign-preserving character at a relatively high $T_\mathrm{c}$ with/without a
significant intraband coupling \cite{Wang2016b,Efremov2016}?

One of the main difficulties in addressing these issues lay in the fact that
superconductivity often occurs in the close vicinity of various competing and
supporting (by their dynamical fluctuations) instabilities, whose microscopic
origins are also not well understood. Another lies in the fact that the degree
of electronic correlations in the FeSCs is not entirely established,  and
likely varies from material to material \cite{Yin2011,deMedici2014}.
Nevertheless, there has been
tremendous progress in modeling the FeSCs, even at a quantitative level using various methods.  

One commonly used approach utilizes Eliashberg-type models, where exchange
bosons often identified with the interband antiferromagnetic spin fluctuations
mediate pairing. Such phenomenological approaches are capable of quantitatively
capturing many experimental observations  
\cite{Johnston2014,Rademaker2016,WangSUST2016,Benfatto2009,Dolgov2013,Ummarino2015,Ummarino2009,Ummarino2013,Ummarino2011}.
The central quantity here is the so-called bosonic (or Eliashberg) spectral
function denoted $\alpha^2F_{ij}({\bf k},{\bf q},\nu)$ in the case of phonons
or $I^2\chi_{ij}({\bf k},{\bf q},\nu)$ in the case of spin-fluctuations.  This
function describes the effective spectrum and its coupling strength for the
exchange bosons that mediating the pairing. (Here, $i,j$ are band indicies, which
allow for inter- and intraband components.) An important quantity is the total
strength of the exchange interaction, which can be quantified 
by integrating the uniform component of the spectral density 
\begin{equation}
\lambda_{ij} = \int_0^\infty \frac{2d\nu}{\nu} \langle\langle \alpha^2F_{ij}({\bf k}, {\bf q},\nu) 
\rangle\rangle,  
\end{equation}
where $\langle\langle \dots \rangle\rangle$ denotes a double Fermi surface
average over the relevant bands.  The band resolved dimensionless couplings
$\lambda_{ij}$ or the band averaged quantity $\lambda_{\rm el-b} =
\sum_{ij}\lambda_{ij}N_i(0)/N(0)$ are of fundamental interest as they can be
related to the average increase in the effective mass of the carriers at the
Fermi level due to the exchange of bosons. [Here, $N_i(0)$ and $N(0)$ denote
the partial and total density of states (DOS) at the Fermi level,
respectively.]  

In this paper, we will review elements of our work supported by the German Research Foundation (Deutsche Forschungsgemeinschaft)
through priority program SPP 1458 and present new high-resolution {\em ab
initio} results for several classes of FeSCs.  
Our goal is to qualitatively discuss several general aspects pertaining to
Eliashberg-based approaches for these materials, which has arisen from these
projects. In another paper in this volume \cite{Efremov2016} we review related
work examining how impurity scattering can induce changes in the symmetry of
the SOCP in such models. Here, we focus on another issue, namely how
measurements of renormalized thermodynamic quantities with respect to their
corresponding
``bare" values can be used to constrain the total electron-boson (el-b)
coupling $\lambda_{\rm el-b}$ in exchange boson scenarios. In doing so, we
place restrictions on semi-empirical Eliashberg-based approaches used to
describe the FeSCs and argue that the boson-exchange scenarios studied to date
place the FeSCs in an intermediate coupling regime, with the band, averaged
coupling $\lambda_{\rm el-b} \stackrel{<}{\sim} 1-1.5$. The physical scenario
for the hole-overdoped AFe$_2$As$_2$
FeSCs (A = K, Rb, Cs) given in the second part of the present paper was
presented at the closing Workshop of the SPP 1458 
hold in 13-17.\ September 2016 in Munich \cite{Drechsler2016}. Details related
to the van Hove singularities will be published elsewhere. 
\vskip 0.5cm 

\subsection{General aspects of el-el interactions.} 
Before we begin, we make some general remarks about the el-el interaction in the FeSCs. 

The phase diagrams of the FeSCs is affected by doping,
pressure, strain, and disorder. But it is unclear which microscopic
interactions drive the relative phases and their boundaries in the phase
diagram. In this context, the proper description of correlation effects in
FeSCs is one of the central and most difficult theoretical problems in the
field. It is becoming more and more apparent, however, that different electron
groups experience varying degrees of correlations, while the overall strength
increases with hole doping \cite{Yin2011,deMedici2014}. One is hence confronted
with a complicated problem of treating two (or even three) electron liquids.
Any attempt to address all of the particles and their interactions
quantitatively and on an equal footing is clearly not possible at present, due
to the extreme complexity of the many-body problem. Nevertheless, analyzing the
superconducting and normal state properties of the FeSCs and comparing them
with band structure 
calculations roughly identifies the Fe 3$d_{xz}$, 3$d_{yz}$, and 3$d_{xy}$
orbitals as crucial for superconductivity, magnetism, and electronically driven
nematicity. 
 
Another related aspect is the impact of the high-energy electronic
interactions, such as the local Hubbard repulsion $U$ and Hund's rule exchange
$J$ on the band structure. These terms are responsible for a significant
overall band narrowing observed across all FeSCs at high energies, which in
turn produces a sizable mass enhancement for the carriers at the Fermi level.
This renormalization is most apparent in the $\sim 2\times$ -- $3\times$
rescaling of the band structure frequently needed to produce a reasonable
agreement between DFT predictions and the bands observed in angle-resolved
photoemission spectroscopy (ARPES) \cite{DMFTReview}, or for thermodynamic
measurements such as the specific heat (see for example Ref.
\cite{Benfatto2009}). Interestingly, the largest mass renormalizations 
have been inferred for CsFe$_2$As$_2$ (Cs-122), which also has one of the
lowest $T_\mathrm{c}$ values. Since the less correlated iron phosphides also
achieve relatively small $T_\mathrm{c}$'s, it is reasonable to infer that
moderate correlations are somehow favorable for superconductivity. In addition
to driving large-scale band renormalizations, the electronic interactions are
also believed to drive pairing in the FeSCs through the formation of
antiferromagnetic spin fluctuations \cite{Mazin2008,ScalapinoRMP} or
possibly other electronic mechanisms \cite{Kontani2010,Kang2016}.  

Electronic correlations can also influence interactions with the lattice, and
it is important to realize that the theory of el-ph coupling in moderately and
strongly correlated materials is not yet fully developed. Indeed, there are
indications that magnetism and electron correlations can significantly enhance
el-ph coupling
\cite{Capone2010,Mandal2014,Coh2015,Coh2016,Sawatzky2009,Drechsler2009,Kulic2009a,Johnston2016a}.
For example, the cations surrounding the Fe-layers can screen the Hubbard $U$
\cite{Sawatzky2009,Drechsler2009,Kulic2009a} and establishing novel lattice
couplings (via modulating the screened el-el interactions \cite{Kulic2009a}, by
changing relevant single particle quantities \cite{Johnston2016a}, {\it etc}.).
Thus, there can be novel sources for the el-ph interactions that are not
accounted for in traditional DFT calculations \cite{Boeri2008}.  These effects
might be responsible for the huge magneto-elastic coupling to the As-derived
A$_{\rm 1g}$ phonon mode in and near the regions of coexistence between
magnetism and superconductivity showing in particular a large deformation
potential of about (0.1 to 0.15) eV/nm in the parent compound BaFe$_2$As$_2$
\cite{Rettig2015}, and the observation of magneto-elastic coupling for
Ba(Fe$_{1-x}$Co$_x$)$_2$As$_2$, even at room temperature \cite{Cantoni2015}
(see also Refs.\ \cite{Barzykin2009,Widom2016,Coh2015}) .  Several observed
phonon anomalies \cite{Kumar2010a,Kumar2010b,Kumar2014} also point to
significant couplings between the lattice and the charge, spin, or orbital
degrees of freedom. Regarding the previously mentioned small empirical values
of $\lambda_{\rm e-ph}$ derived from the high-energy relaxation rates, we
remind the reader that this measurement accesses the transport coupling
constant, which weights backscattering processes and differs from el-ph
coupling that enters Eliashberg theory \cite{Zeyher1996}. These considerations
show that el-ph coupling cannot be ruled out {\it a priori} in moderately to
strongly correlated materials based on DFT arguments alone. Indeed, some
studies have inferred a significant contribution from the lattice
\cite{Ummarino2015,Coh2015,Coh2016,Rademaker2016,Johnston2014}. 
\vskip 0.5cm 

\subsection{Mass renormalizations and band shifts.}
Other fundamental and still unresolved questions concern the role of
retardation (frequency) and the particular momentum dependence of the
interactions responsible for pairing (dominated for instance by specific
nesting vectors \cite{Mazin2008} for small momentum transfers
\cite{Lee2014,Rademaker2016,Kang2016}) .  A full accounting of all of these
effects in detail is practically impossible for the multi-band Fermi surfaces
present in the FeSC; nevertheless, the effective mass $m^{*}$ of the
quasiparticles near the Fermi level of a Fermi-liquid-like system can provide
some insight into the overall strength of the many-body effects and the
relevant interactions.  This information is encoded in the real part of the
complex-valued electron  self-energy $\Sigma({\bf k},\omega) =
\Sigma^{\prime}({\bf k},\omega) + {\rm i}\Sigma^{\prime\prime}({\bf
k},\omega)$,
\be
\frac{m^*(\vec{k}_{\rm F})}{m_{\rm \tiny b}(\vec{k}_{\rm F})}
=\frac{ 1 -\partial \Sigma^{\prime}
\left( \omega , T, \vec{k}_\mathrm{F} \right) /\partial \omega }
{ 1-\frac{\partial \Sigma^{\prime}
\left( \omega, T , \vec{k}_\mathrm{F}  \right) }{\partial \vec{k}   } }, \quad
\label{mass1}
\ee
where $m_{\rm \tiny b}$ is the bare mass in the absence of interactions and
$\omega$ is the quasiparticle energy measured relative to the Fermi level.
\footnote{The self-energy can also result in significant shifts of the
individual bands, which can dramatically change the effective mass in cases
where empty bands are near the Fermi level, or in the presence of shifted van
Hove singularities . These effects can be necessary for the understanding both
the normal and superconducting states of the FeSCs and are discussed in
Sec.~\ref{Sec:Shifted}.}

For our purpose, we will take $m_\mathrm{b}$ as the band mass computed by DFT
calculations, where the correlation effects introduced by the various DFT
exchange-correlation potentials (Slater, Perdew, etc.) are small.  In doing so,
the ratio $m^*/m_\mathrm{b}$ then gives an approximate measure of the strength
of the interactions beyond the single-particle picture provided by DFT. 

As mentioned, the electronic interactions in the FeSCs provide a contribution
to the total mass enhancement at the Fermi level. However, they are also
believed to be responsible for the exchange bosons that act as the pairing
mediator in these materials.  According to most scenarios, the exchange bosons
are restricted to a low energy interval of approximately 0 -- 300~meV,
considerably below the typical energy scale of the Hubbard $U\sim 2-4$ eV or
Hund's exchange $J \sim U/4$ \cite{Maier2008}. Thus, in these frameworks, the
pairing interaction represents only a fraction of all interactions present in
the system. Similarly, the el-b interaction must also provide only a portion of
the total mass $m^*$ or the relative energy shifts of the various bands.
[Formally, one can extend such bosonic descriptions to higher energies
\cite{Iwasawa2010,Iwasawa2013} using flexible model expressions for the
self-energy entering Eq.\ (\ref{mass1}), and such approaches have been applied
successfully to Sr$_2$RuO$_4$ and other 4$d$ systems.] Based on this, we
suppose that the total coupling $\lambda$ can be partitioned into ``high"-
and ``low"-energy contributions, where the typical energies differ at least by
an order of magnitude. Here, we regard the high-energy part as encompassing
energies ranging from about $500$ meV to the bandwidth, while energies $\le
200$ -- $500$ meV represent the residual bosonic excitations entering the
kernels of the Eliashberg equations. The latter in principle contains the
action of the phonons as well as any spin and charge fluctuations established
by the electronic interactions. (Note that this is an approximate partition. In
the vicinity of an orbital-selective Mott transition, the Hund's rule coupling
$J$ can also be active at low energies and high temperatures through its role
in  determining the ``bad metal" incoherence regime.)

With this partition in mind, it is convenient to rewrite Eq.\ (\ref{mass1}) in
an approximate factorized form, which separates the contributions from the two
energy regimes 
\begin{eqnarray}
\frac{m_*}{m_\mathrm{b}}&=&(1+\lambda) \\ \nonumber
&\approx&
\left(1+\lambda^{\rm high}_{\rm el-el}\right)
\left(1+\lambda^{\rm low}_{\rm el-b}(T)\right) \\ \nonumber 
&\approx&\frac{m^*_{\rm high}}{m_b}\left(1+\lambda^{\rm low}_{\rm el-b}(T)\right). 
\end{eqnarray}
Here, $\lambda$ denotes to {\em total} coupling strength including the effects
of all interactions, $\lambda^{\rm high}_{\rm el-el}$ is the Fermi-surface
averaged contribution of the el-el interactions at high-energies and
$\lambda^{\rm low}_{\rm el-b}(T)$ is the low-energy contribution (renormalized
by the pre-factor \cite{Iwasawa2013}) from the various retarded interactions
mediated by exchange bosons.  (For brevity we will drop the $T$ dependence from
our notation;  however, this quantity can vary with $T$ when superconductivity
is fed back into the bosonic spectrum.) The latter includes the bosons acting
as the superconducting glue and potentially others that do not contribute to
pairing but still dress the quasiparticles. For instance, in $d$-wave
superconductors  the latter might include bosons with a momentum-independent
coupling, which will contribute the effective mass while providing no
contribution to $d$-wave pairing \cite{Bulut1996,JohnstonDOS}. 
\newline

\subsection{Quantifying the high-energy renormalization.}
A rough estimate of the high-energy contribution to the mass renormalization
can be obtained by comparing measurements of quantities such as the total
electronic bandwidth or plasma frequency against the predictions of DFT. For
example, the value of  $1+\lambda^{\rm high}_{\rm el-el}$ should be
approximately equal to the ratio of the overall measured and DFT-calculated
electronic bandwidths.  Similarly, ignoring some uncertainties related to
impurity scattering, one can get a similar estimate by comparing the squares of
the calculated total intraband plasma frequency $\Omega^2_{\rm p} = \sum_i
\Omega^2_{{\rm p},i}$, where $i$ is a band index, to the experimental values
for $\Omega^2_{\rm p}$ at room temperature. 

We have carried out such an exercise for several of the FeSCs using the total
plasma frequencies, and compare the experimental plasma frequencies to the
values obtained from our DFT calculations in Tbl.~1. For La-1111 or Ba-122 we
find a value of $1+\lambda^{\rm high}_{\rm el-el} \sim 2.78$,  while for the
more correlated K-122 we find $1+\lambda^{\rm high}_{\rm el-el} \sim 6$. 

\begin{table}
\begin{tabular}{ccccc}
 \hline
  Compound & $\Omega_{\rm p}$ (DFT) & $\Omega_{\rm p}$ (Exp.) & $\frac{m^{*}_{\rm high}}{m_{\rm b}}$ &  \\
  \hline
  LaOFeAs &2.2 & 1.33 &2.7$\pm 0.1$& P \\
LaO$_{0.9}$F$_{0.1}$FeAs &2.2& 1.37& 2.33& P\\
LaO$_{0.9}$F$_{0.1}$FeAs$_{0.9}$ & & $\leq 1.3$ & &P\\
SrFe$_2$As$_2$ &2.8 &1.72 &2.63 &S\\
La$_x$Sr$_{1-x}$Fe$_2$As$_2$& 2.8&&&  \\
SrFe$_{2-x}$Co$_x$As$_2$ &2.7&&& \\
BaFe$_2$As$_2$& 2.63& 1.66& 2.51& S\\
BaFe$_2$As$_2$&& 1.58& 2.78& S\\
K$_x$Ba$_{1-x}$Fe$_2$As$_2$& 2.63& 1.6& 2.78& S \\
K$_{0.45}$Ba$_{0.55}$ Fe$_2$As$_2$ &2.63 &$1.7 $ &2.39 &S\\
CaFe$_2$As$_2$ &2.95&1.85&2.56&S\\
CaFe$_2$As$_2$ &2.95&2.71&1.19&S\\
CaFe$_2$As$_2$ &2.95&2.33&1.6&S\\
LiFeAs &2.9&1.93&2.27&S\\
KFe$_2$As$_2$&2.54&1.04&6.0&S\\
\hline
\end{tabular}
\caption{Calculated (DFT) and experimental unscreened in-plane
plasma frequencies ($\parallel $ ab) in units of eV. 
We used the virtual crystal approximation for the calculations of the doped
systems. P and S stand for polycrystalline sample and single crystal,
respectively. The experimental data are taken from
Ref.\ \cite{Dai2016} for LiFeAs,
Ref.\ \cite{Cheng2012} for CaFe$_2$As$_2$.  For the
other compounds see, Ref.\ \cite{Drechsler2010} \label{tab.1}.
Here, $m_{\rm b}$ is obtained from DFT calculations.
 }
\end{table}

Before proceeding, we note that in the case of CaFe$_2$As$_2$ we have adjusted
the values of $\Omega_{\rm p}$ from the ones reported in Ref. \cite{Cheng2012}.
Typically one obtains the plasma frequency in the FeSCs by fitting optical
conductivity data with at least {\it two} Drude peaks for the sake of
simplicity. In doing so, one usually finds (see e.g.\ Ref.\
\cite{Maksimov2011}) that one of the Drude peaks is quite broad compared to the
other.  This also happens in the two special cases under consideration Ca-122
and K-122 \cite{Cheng2012,Wu2010,Charnukha2013,Dai2016a}. This simple and
minimal fitting procedure, which is at first glance reasonable, might yield
misleading results, however. This possibility is evident in the fact that it
sometimes results in $\Omega_{\rm p}$ values that are comparable to or exceed
the calculated DFT values, leaving no room for additional renormalizations by
interactions.  For example, Ref.\ \cite{Cheng2012} reported a $\Omega_{{\rm
p},2}=20500$~cm$^{-1}$ (2.54~eV), which nearly coincides with our calculated
total $\Omega_{\rm p}$, and yields in turn a small mass renormalization and a
large scattering rate corresponding to a mean free path $l < 10$~nm. In our
opinion, the solution to this puzzle is to decompose the unusually broad Drude
peak into two subcomponents: a narrower Drude peak with a smaller partial
$\Omega_{\rm p}$ and an additional interband transition at relatively low
energies.  

We believe that this is a general phenomenon in the FeSCs, where the broad
Drude peak frequently obtained from optics data likely contains a contribution
from a low-lying interband transition (probably below the first Lorentz peak at
about 6000~cm$^{-1}$ [0.75~eV]). This interband transition is typically
unresolved due to the nearly structureless optical spectra below 1.2~eV (shown
for example in Fig.\ 5b of Ref.\ \cite{Cheng2012}). This hypothesis is
supported by reports of such interband transitions in measurements for Ba-122
\cite{Tytarenko2015,vanHeumenEPL} and in DFT calculations
\cite{BenfattoInterband}. It can further be corroborated using an alternative
measure of $\Omega_{\rm p}$ obtained by integrating the real part of the
dielectric function \cite{Cheng2012}. In the case of Ca-122, one would arrive
at $\Omega_{\rm p} \approx 1.96$~eV and a mass renormalization of 1.68, which
is clearly still too small.  But for the weakly correlated undoped Ca-122
\cite{Diel2014}, this procedure yields a more reasonable mass renormalization
of 2.56 when one adopts 15000~cm$^{-1}$ (1.86 eV) for the experimental
$\Omega_{\rm p}$. This value is a bit smaller than one of the available data
for Ba-122 with larger lattice constants. For the heavily hole doped
Ca$_{0.32}$Na$_{0.68}$Fe$_2$As$_2$ \cite{Johnston2014} a markedly larger mass
renormalization of 4.25 $\pm 0.25$ is obtained using the corresponding
$\mu$SR-data \cite{Materne2015} and employing the relations given in Refs.\
\cite{Drechsler2009,Drechsler2008}. For this reason, in the case of K-122 being
of special interest (see Sec.\ 3), we replaced the experimental $\Omega_{{\rm
p},2}=20500$~cm$^{-1}$ \cite{Cheng2012} by a much smaller value of
$\tilde{\Omega}_{\rm p2}=7514~$cm$^{-1}$ (0.93~eV).  The latter exceeds the
experimentally well-defined lower $\Omega_{{\rm p},1}=6500$~cm$^{-1}$ (0.81~eV)
obtained for the other Drude component. As a result, we arrive at a
dominant high-energy mass renormalization of about six necessary to describe
its large Sommerfeld constant $\gamma$ and the superconductivity mediated by
weaker low-energy bosonic interactions (see below).

\begin{figure}[t]
\centerline{\includegraphics[width=0.75\columnwidth]{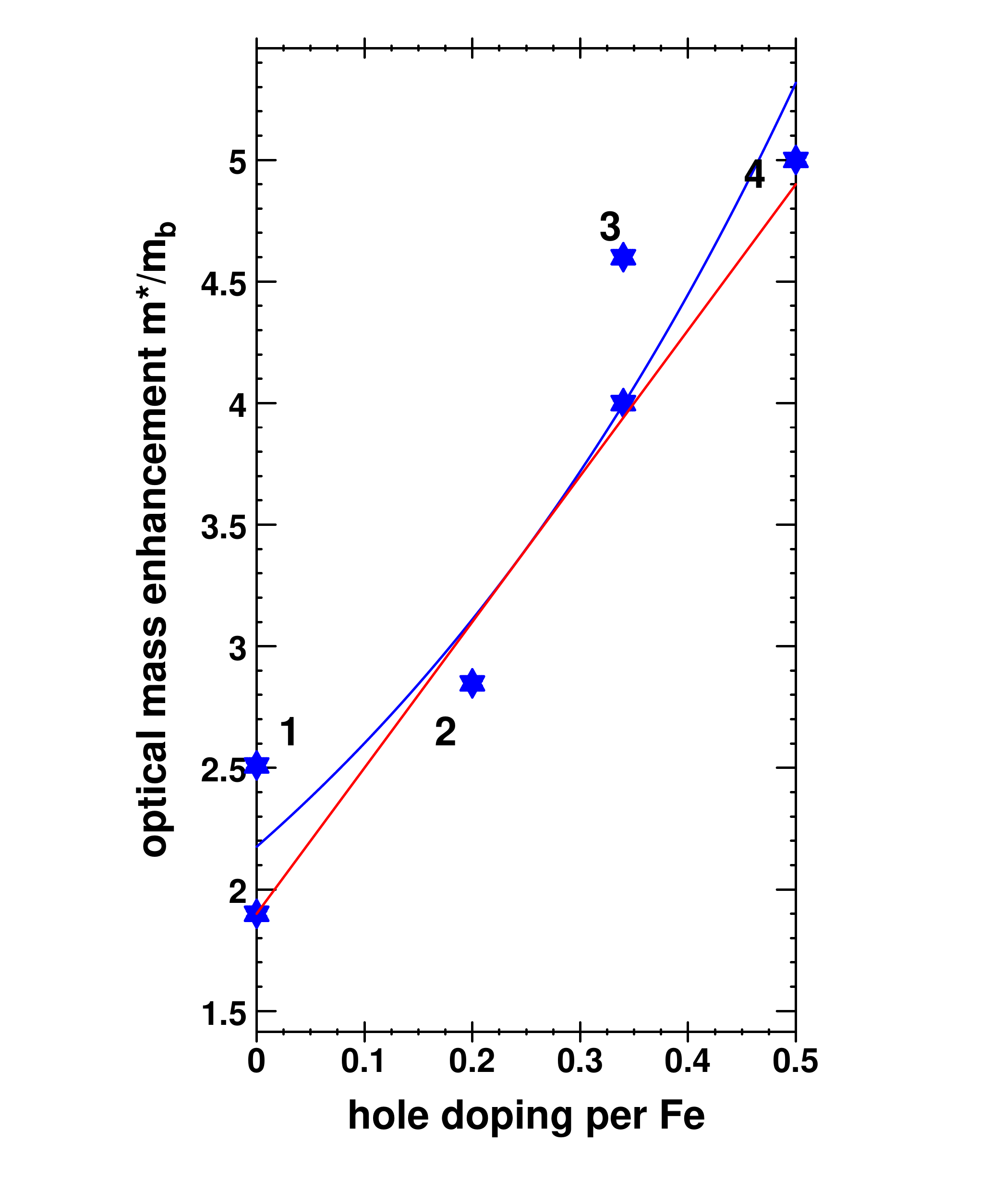}}
\caption{({color online}) The empirical optical mass enhancement
for various relative clean 122 FeSCs with doping outside the FeAs- layers
and different hole content using the data shown in  Tab.\ 1:
Red and blue curves are linear and exponential regression based
guides to the eyes, respectively. The large error bars in the optical data due the presence
of several interband transitions do not allow to resolve the expected positive curvature.}
 \label{massopt}
\end{figure}

All of the above analysis has been carried out using the room temperature
plasma frequency in the normal state. If one wishes to do a similar study in
the superconducting state, then some additional care must be taken to
account for the possibility that not all electrons take part in forming the
superconducting condensate. 
For example, a crude rescaling of the zero temperature London penetration depth
$\Lambda$ for hole-overdoped system Ba$_{0.35}$Rb$_{0.65}$Fe$_2$As$_2$
\cite{Guguchia2016}, which is a candidate for $d$-wave superconductor as in
Rb-122, gives     
\begin{equation*}
 \frac{\Omega_{\rm p}^{\mbox{\rm \tiny BaRb}}}{\Omega_{\rm p}^{\mbox{\tiny CaNa}}}
 \frac{ \Lambda^{\mbox{\tiny BaRb}}(0)}{\Lambda^{\mbox{\rm \tiny CaNa}}(0)}
 \approx
\sqrt{
\frac{(1+\lambda^{\mbox{\tiny BaRb}})n^{\mbox{\tiny BaRb} } n_{\rm s}^{\mbox{\tiny CaNa} }}
     {n_{\rm s}^{\mbox{\tiny BaRb}} n^{\mbox{\tiny CaNa}}(1+\lambda^{\mbox{\tiny CaNa}})}
},
\end{equation*}
where $n$ is the total charge density of all conduction electrons, and $n_s$ is
the part residing in the condensate at $T=0$. Note that for an electronic or
magnetic pairing mechanism, some of the electrons are components of the pairing
interaction, and therefore one expects $n_s/n < 1$. \footnote{This effect is
usually small and was ignored in Ref.\ \cite{Johnston2014}, were we adopted
$n_s/n \approx 1$ for the optimally Na-doped Ca-122 system for the sake of
simplicity. Below we will consider a much more dramatic case, where one band
with a substantial DOS and a marked contribution to the total plasma frequency
is quenched at low temperature.} Using the experimental London penetration
depth data of $\Lambda^{\mbox{\tiny BaRb}}=257$~nm \cite{Guguchia2016}
$\Lambda^{\mbox{\tiny BaRb}}=194$~nm \cite{Materne2015}, $\lambda^{\mbox{\tiny
CaNa}}=0.89$ \cite{Johnston2014}, and adopting a comparable total coupling
constant $\lambda^{\mbox{\tiny BaRb}}\approx 1$ and   similar unscreened plasma
frequencies, we estimate $n_{\rm s}^{\mbox{ \tiny BaRb}}\approx 0.6$, only.
This way by quenching the weakly coupled $d_{xy}$-derived band by the combined
effect of a weak magnetic field and disorder, the single-gap $d$-wave scenario
might be in principle understood.

We have performed a similar phenomenological analysis of different optical
conductivity and penetration depth measurements for several FeSCs as a function
of hole doping. The results, summarized in Fig.\ \ref{massopt}, reveal a steep
increase of the mass renormalization with hole doping, in qualitative agreement
with the Hund's metal picture proposed by the DMFT \cite{HundsMetal} and the
slave-boson approximations \cite{Medici2016}.  Similar constraints on the value
of $\lambda^{\rm high}_{\rm el-el}$ can be placed on other compounds. For
example, an orbital-resolved ARPES analysis for the undoped LiFeAs and the
el-doped Ba(Fe$_{0.92}$Co$_{0.08}$)$_2$As$_2$ results in comparable numbers for
the band renormalization renormalization: about 2.3 and 2.1 for the el-pockets
with 3$d_{xz}/d_{yz}$ and 3$d_{xy}$ character in Ba-122 and 2.2 (1.8) for the
el (h)-pockets with 3$d_{xz}/d_{yz}$ character, but 4 (3.3) for the el
(h)-pockets  of 3$d_{xy}$ character in LiFeAs. The increased values of the
former are in accord with the DMFT predictions for the importance of
correlation effects for the 3$d_{xy}$ states with further h-doping.
\cite{Brouet2016,Roekeghem2016,Guterding2016,Backes2015}.

\section{Weak versus strong coupling in Fe-based superconductors.}
Having obtained rough estimates for the high-energy contribution to the mass
renormalization, we now turn our attention to the low-energy contribution
coming from el-b interactions. 

The single-band BCS model describes many weak-coupling superconductors.  This
model predicts universal thermodynamic relations in the weak coupling limit.
While this is consistent with measurements on weakly coupled conventional
superconductors, notable deviations occur in strongly-coupled cases.
Strong-coupling Eliashberg theory can account for these discrepancies by
taking into account the retarded nature of the effective attractive
interaction mediated by phonons \cite{Carbotte1990}. Based on this success,
many have attempted to generalize this approach to unconventional
superconductors, which are believed to be strongly coupled due to their large
values of $T_\mathrm{c}$. In the case of the FeSCs, such schemes must also be
generalized further to include multiband effects. Implicit in this approach is
the assumption that boson-exchange frameworks are the appropriate language to
describe pairing in unconventional superconductors
\cite{Maier2008,Anderson2007}. Here, we will not discuss the validity of this
outlook but merely comment on the restrictions placed on it in the context of
the FeSCs. 

To estimate the total interaction with low-energy bosonic modes, we must
examine quantities that are sensitive to the mass changes at the Fermi level.
One such quantity is the Sommerfeld coefficient $\gamma$, which characterizes
the linear in $T$ contribution to the electronic specific heat. For an
interacting system $\gamma$ is written as 
\be
\gamma=\left( 1+\lambda\right) \gamma_\mathrm{b}, 
\quad \quad \gamma_\mathrm{b}=\frac{2}{3}\pi^2k_{\rm B}^2N_\mathrm{b}(0),
\ee
where $\gamma_\mathrm{b}$ is the ``bare" Sommerfeld coefficient for the
non-interacting system, $k_{\rm B}$ is the Boltzmann constant,
$N_\mathrm{b}(0)$ is the bare DOS at the Fermi level (taken from band structure
calculations), and $\lambda$ is the {\em total} band-averaged coupling constant
due to all interactions. Therefore, if one knows $\gamma$ and $\lambda_{\rm
el-b}^{\rm high}$, one can estimate $\lambda_{\rm el-b}^{\rm low}$. 

The values of the average and band-resolved coupling constants at low-energy
$\lambda_{\rm el-b}^{\rm low}$ in the FeSCs are still under debate, and several
Eliashberg-based studies arrive at different values for the total coupling.
Shortly after the discovery of FeSC, LaFeAsF$_{0.1}$O$_{0.9}$ was identified as
being in an {\it intermediate} coupling regime with $\lambda_{\rm el-b}^{\rm
low}=0.61\pm 0.35 < 1$ \cite{Drechsler2008,Drechsler2009,Drechsler2010}.  This
estimate was obtained using the well-known mass dependence of the
superconducting condensate density determined by the low-temperature
penetration depth measurements. Similarly, the upper critical field $H_{c2}(T)$
for a sample with strong paramagnetic effects due to the presence of As
vacancies were also successfully described using the same weak to intermediate
coupling regime \cite{Fuchs2008,Fuchs2009}.  In contrast, a $\lambda_{\rm
el-b}^{\rm low} \sim 2$ was obtained for optimally doped
Ba$_{1-x}$K$_x$Fe$_2$As$_2$ ($T_\mathrm{c} = 38$ K)
\cite{Charnukha2011,Popovich2010} from specific heat and optical data within a
four-band Eliashberg model. This result is larger than the $\lambda_{\rm
el-b}^{\rm low} < 1.5$ estimated by L.\ Benfatto {\it et al.}
\cite{Benfatto2009} for the same compound also within a four-band model. 

\begin{figure}[b]
\centerline{\includegraphics[width=0.99\columnwidth]{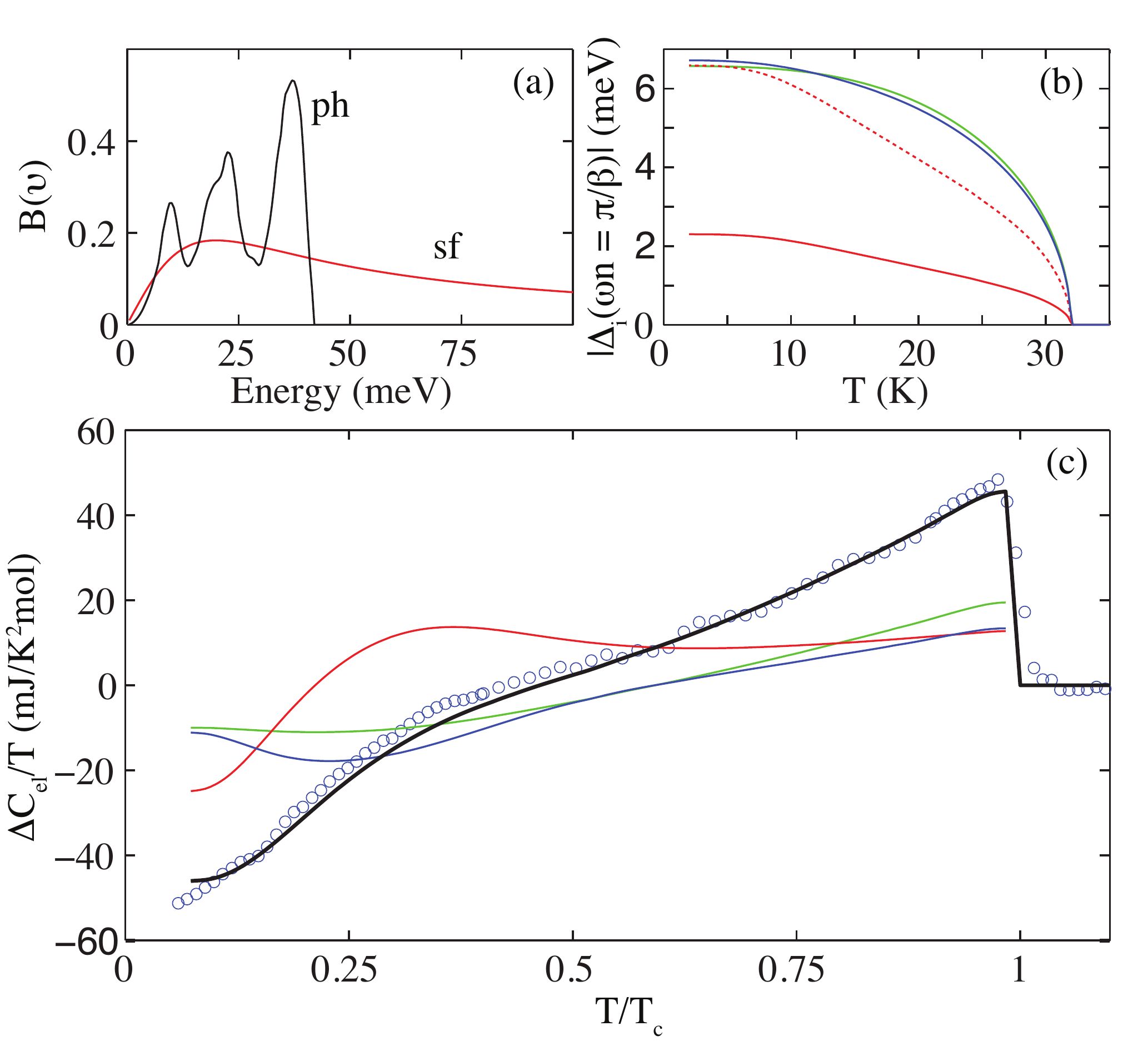}}
\caption{({color online}) Adopted Eliashberg-functions (a),
calculated gaps (b) and the fitted
electronic specific heat (c)
employed for the theoretical description of optimally Na-doped Ca-122 (see
text) as described within an effective three-band model, taken from Ref.
\cite{Johnston2014}. Notice the hump at about 0.35$T_\mathrm{c}$ which stems
from the phenomenologically
large PDOS of the weakly coupled third band with dominant Fe 3$d_{xy}$
character reflecting this way the high-energy mass renormalization. A similar
feature is observed at weak interband coupling  for the weakly coupled
$\beta$-band in K(Rb,Cs)-122 (see Sect.\ 5)}.
 \label{Fig:Ca122}
\end{figure}

We recently performed similar calculations for the specific heat of
Ca$_{0.32}$Na$_{0.68}$Fe$_2$As$_2$ ($T_\mathrm{c}=34.4$~K) using a three-band
model. Figure \ref{Fig:Ca122} reproduces the main results. By fitting the data
we arrived at $\lambda_{\rm el-b}=0.88$ \cite{Johnston2014}.  The three-band
description allowed us to decompose the involved coupling constants, which
included a dominant pair of bands coupled by a significant repulsive interband
coupling with $\lambda_{12}=-1$ and a weakly coupled third band with
$\lambda_{13}=-0.1$ and $\lambda_{23} = 0$. Surprisingly, we also obtained a
non-negligible attractive intraband interaction (interpreted as due to el-ph
coupling or perhaps orbital fluctuations)
$\lambda_{11}=\lambda_{22}=\lambda_{33}=0.45$ (taken to be equal in this case
to reduce the number of fitting parameters).  This value is twice as large as
the typical DFT based estimates \cite{Boeri2008} but was needed to reproduce
the pronounced knee observed in the specific heat at $T/T_\mathrm{c}\sim0.3$.
In this model, the repulsive interband interaction still provides the majority
of the total $T_\mathrm{c}$: switching off the el-ph interaction, we obtained a
relatively high $T_\mathrm{c}=21.7$~K, but switching off the spin-fluctuations
produced $T_\mathrm{c}=5.4$~K. 

Our model for Ca$_{0.32}$Na$_{0.68}$Fe$_2$As$_2$ predicted $T=0$ gap values of
$\Delta_1=7.48$~meV, $\Delta_2=2.35$~meV, and $\Delta_3=7.5$~meV, in agreement
with ARPES \cite{Evtushinsky2013} ($\Delta_1\approx \Delta_3=7.8$~meV and
$\Delta_2=7.48$~meV) and $\mu$SR (6 -- 6.7~meV and 0.6 -- 0.8~meV within a
finite magnetic field) measurements.  Another ARPES study \cite{Shi2014}
arrived at somewhat larger gaps \cite{Shi2014} for a single crystal with the
same $T_\mathrm{c}$ (see Tab.\ 2). Experimentally, the smallest gap occurs on
the outermost hole Fermi surface with $d_{xy}$-character. 

 \begin{table}[b]
\begin{tabular}{ccccc}
 \hline
  $\Delta_i(0)$& ARPES & ARPES & $\mu$SR & C($T$)  \\
  \hline
 Ca$_{0.33}$Na$_{0.67}$ Fe$_2$As$_2$ & & & &\\
$\Delta_1$ & 10.2 & 7.8 & 6.7 $\pm 1.3$ &  7.5\\
$\Delta_2$ & 5.7 & 2.3 & 0.7$\pm 0.1$ & 2.35\\
$\Delta_3$&9.2&7.8&6.7$\pm 1.3$&7.48\\
\hline
Ba$_{0.33}$Rb$_{0.65}$Fe$_2$As$_2$& & & &  \\
$\Delta_1$ & & & 8.4& \\
Ba$_{0.1}$K$_{0.9}$ Fe$_2$As$_2$ & & & &\\
$\Delta_1$ & 3.2 &--  &--  & -- \\
$\Delta_2$ & 2.9 &--  &--  & --\\
$\Delta_3$& 2.7 & --&--&--\\

\hline
\end{tabular}
\caption{Experimental and theoretical values for multiband gaps in meV
of two strongly hole doped 122 FeSC,Na-doped Ca-122 and Rb-doped Ba122,
described in effective single, two-band and three-band models as derived from
ARPES, specific heat and penetration depth ($\mu$SR) data (see text).
 }
 \label{tab2}
\end{table} 

Several independent analyses have arrived at similar estimates for
$\lambda_{\rm el-b}^{\rm low}$.  For example, a recent analysis of optical data
for nearly optimally doped Ba$_{0.6}$K$_{0.4}$Fe$_{2}$As$_{2}$
($T_\mathrm{c}=39$~K) and LiFeAs ($T_\mathrm{c}=17$~K) by J.\ Hwang
\cite{Hwang2016} extracted the el-boson spectral densities (the Eliashberg
function) within an effective single band approach.  The measured total plasma
frequencies of the superconducting condensate $\Omega^2_\mathrm{c}\approx
\Omega^2_{\mbox{\tiny unscreened}}/(1+\lambda^{\rm low}_{\mathrm el-b})$ are
1.01~eV for Ba$_{0.6}$K$_{0.4}$Fe$_{2}$As$_{2}$ and 0.87~eV for LiFeAs. In the
clean limit, this yields $\lambda^{\rm low}_{\rm el-b}=1.51$ for the former (if
the empirical value of $\Omega_{\mbox{\tiny unscreened}} = 1.6$~eV
\cite{Charnukha2011} is used instead of the {\it adopted } 1.8~eV and claimed
$\lambda^{\rm low}_{\rm el-b}=1.98$ by the author).  This value is in perfect
agreement with our estimate based on the Sommerfeld constant $\gamma$ and a
high-energy renormalization of 2.5 inferred from the ARPES data
\cite{Derondeau2016}.  This value also agrees well with the four-band analysis
by L.\ Benfatto {\it et al.} \cite{Benfatto2009}, and is also closer to 0.9
obtained for optimally Na-doped Ca-122 \cite{Johnston2014}.  Thus, three
independent analysis arrive at an intermediate coupling regime for FeSCs with
$T_\mathrm{c}\le 40~K$. A weak coupling regime is also implied for LiFeAs.
Using the recent optical data for LiFeAs \cite{Dai2016} (see also Tab.\ I) with
$\Omega_{\mbox{\tiny unscreened}}= 1.93$~eV.  Similarily, several analyses for
LiFeAs have arrived at small values for $\lambda^{\rm low}_{\rm el-b}$. For
example, an analysis of specific heat similar to that shown in Fig.~2
\cite{Johnston2016} arrives at $\lambda^{\rm low}_{\rm el-b} \approx 0.6$ while
comparable values of $\lambda^{\rm low}_{\rm el-b} \approx 0.8$ where obtained
from scanning tunneling microscopy  \cite{Hlobil2016,Schmalianprivate2016} and
0.89 from  optical data \cite{Hwang2016} (once a realistic value of the
unscreened plasma frequency is adopted).
 
The possibility of changing the coupling regime from weak to strong across the
various Fe pnictide families was proposed in Ref.\ \cite{Inosov2011}, based on
an analysis of the $2\Delta(0)/k_\mathrm{B}T_\mathrm{c}$ ratio, as well as the
jump in the specific heat $\Delta C_v(T_\mathrm{c})/T_\mathrm{c}$.  According
to that analysis, optimally K-doped Ba-122 is the most strongly coupled FeSC.
However, given the collection of results summarized above, we believe it is
clear that many FeSCs have an intermediate coupling to the low-energy bosonic
modes. Furthermore, we are in a position to explain several difficulties in the
$\lambda^{\rm low}_\mathrm{el-b} \sim 2$ estimated obtained for
Ba$_{1-x}$K$_x$Fe$_2$As$_2$, which underestimates the high-energy
renormalization.  Using the experimental value of the Sommerfeld coefficient
$\gamma \approx 50$~mJ/mol $\cdot $K$^2$ \cite{Kant2010} and the DFT-calculated
$\gamma_b = 9.26$~mJ/mol $\cdot $K$^2$ \cite{Ma2010}, we arrive at $(1+\lambda)
= \gamma/\gamma_b \approx 5.3$, which must be partitioned between the high- and
low-energy interactions.  The high-energy contribution can be estimated as
outlined in the previous section.  Ref.\ \cite{Derondeau2016} reported
significant high-energy bandwidth renormalizations of 2.94 and 1.98 for the
inner and outer hole pockets at $\Gamma$, and 1.58 for the hole pocket at the
$X$-point. If we adopt comparable PDOS-values for each band, we estimate
$1+\lambda_{\rm el-el}^{\rm high} < 2.2$, which is a little lower than the
values $\approx 2.4$ -- $2.7$ obtained from the plasma frequencies (Tlb. 1).
From these estimates, we then see that $\lambda^{\rm low}_\mathrm{el-b} \sim 2$
would require a $\lambda^{\rm high}_\mathrm{el-el} \sim 1.77$, which is too
small compared to any of our estimates.  By comparison, $\lambda^{\rm
low}_\mathrm{el-b} \le 1.5$ allows for a $1+\lambda_{\rm el-el}^{\rm high} \sim
2.2$, which is within our range. 

This discrepancy likely stems from the use of the DFT-derived PDOS in Ref.
\cite{Popovich2010} in place of the high-energy renormalized values (as
considered in Ref. \cite{Benfatto2009}).  Another factor might be related to
some underestimation of the strength of the el-ph (or orbital fluctuation)
interaction adopted in the modeling of the multiband interaction matrix.  A
similar overestimation of the spin-fluctuation interaction is present in the
model calculations by Ummarino {\it et al.}
\cite{Ummarino2009,Ummarino2013,Tortello2010}.  There, the low-energy bosonic
coupling constants are $\lambda_{\rm el-b}^{\rm low} \sim 4 - 5$, which has the
same fatal consequences for the Sommerfeld constant $\gamma$ and the
high-energy mass renormalizations. Only in their most recent paper (devoted
mainly to FeTe$_{1-x}$Se$_x$ \cite{Ummarino2015} analyzed within a three-band
model), did the authors allow for a sizable intraband coupling ascribed to the
el-ph interaction present in the third el-pocket, only.  This way they arrive
at $\lambda_{\rm el-b}^{\rm low}=1.48$ and $\lambda^{\rm low}_{\rm el-ph}=1.1$.
The last value seems to be somewhat large, but it can be lowered by adopting a
weak intraband coupling in the remaining two bands as well.  It is also
noteworthy that the authors demonstrated the significance of a self-consistent
calculation by accounting for the feedback of superconductivity on the spin
fluctuations within their approach, which is usually ignored in Eliashberg
calculations. 

For the remainder of this section we will discuss several subtle issues that
can affect assessments of the total el-b coupling in the FeSCs.  In this
context, two more results of Ref.\ \cite{Hwang2016} are noteworthy.  First, in
addition to the dominant narrow peak near 15~meV expected for the interband
spin fluctuations, the extracted spectral density has a second broader peak
centered at $\sim 48$~meV.  This higher-energy peak is absent in the models
adopted in Refs.\ \cite{Benfatto2009,Charnukha2011,Popovich2010}.  Second, the
analysis of the normal state optical properties at 40~K (just above
$T_\mathrm{c}$) shows a hardening of the first dominant peak compared with the
data deep in the superconducting state ($T=4$~K). This observation yields a
$\lambda^{\rm low}_{\rm el-b}$ that is enhanced by about 10\% as compared to
the normal state.  Such an empirical $T$-dependent boson spectrum was not
accounted for in Refs.\
\cite{Johnston2014,Benfatto2009,Charnukha2011,Popovich2010} and might have an
impact on the temperature dependence of the specific heat and the zero
temperature gap values. Verifying the 
$T$-dependent spectral density by other measurements is of considerable
interest since it provides new insight into the interplay between magnetism and
superconductivity. 

Ignoring the high-energy band renormalizations can result in inaccurate estimates 
of 
the size of the of specific heat jump $\Delta C/\gamma_n T_\mathrm{c}$ and other observables.  
Another 
many-body effect that is potentially significant for multiband systems is the
possible chemical potential shifts associated with charge redistribution below
$T_\mathrm{c}$ \cite{Ortenzi2009,Benfatto2011}.  Any strong coupling regime
deduced from a simple Eliashberg-theory calculation that ignores changes of the
chemical potential in the superconducting state -- especially for narrow or
shallow bands -- will miss this contribution to the jump in the specific heat
and other observables. As an instructive illustration, we refer to recent work
using simpler models where such chemical potential shifts were neglected
\cite{Linscheid2016,Chubukov2016,Valentinis2016}.  An advanced multiband
Eliashberg analysis along these lines is necessary to reconsider/check the
strong coupling results mentioned above.

A weak-coupling interband regime within two- and three-band descriptions has
also been found for Co-doped Ba-122 \cite{Maksimov2011,Karakozov2014}, based on
specific heat and optical data. These works found a non-BCS shape of the
$T$-dependent small gap functions, which affected the behavior of the
penetration depth, in contrast to a BCS-like shape expected for a dominant
interband coupling.  This shape clearly points to the presence of a significant
intraband coupling in this el-doped system, similar to Fig.\ \ref{Fig:Ca122}.
It also might explain the possibility for a symmetry change of the SCOP with
increasing disorder at a reduced but still significant $T_\mathrm{c}$
\cite{Efremov2016}. 

Finally, let us consider the case of strongly el-doped LaFeAsO$_{1-x}$H$_x$ and
$x=0.2$, where a rather large $T_\mathrm{c}=48$~K (52~K) was reported under
3.0~GPa (6.0~GPa) of pressure  \cite{Kawaguchi2016}. The increase in $T_c$ was
observed   {\it without} any indication corresponding to stronger spin
fluctuations (which might at first glance explain the significant $T_c$
enhancement under pressure) as deduced from both NMR and inelastic neutron
scattering measurements.  In such a situation a dominant conventional intraband
mechanism based on phonons or orbital/charge fluctuations is very likely.  A
rough estimate for the corresponding coupling constant can be obtained from an
approximate analytical expression for $T_\mathrm{c}$ (derived for an
intermediately coupled effective single-band superconductor
\cite{Maksimov1982}) 
\begin{equation}
\lambda=\frac{1}{\left[ \ln \left( \omega_{\rm ln}/k_{\rm B}T_\mathrm{c} \right)-1-A+ \ln(1.13) \right]}. 
\label{Hosono}
\end{equation}
Adopting $\omega_{\rm ln} \sim 35$ -- $40$~meV, as suggested by couplings to
As-phonons, and a shape factor $A \approx 0.5$ valid for a narrow peak in the
Eliashberg function, one arrives again at $\lambda^{\rm low}_{\rm el-b} = 1.32$
(1.48) for $\omega_{\rm ln} = 35$ meV and $\lambda^{\rm low}_{\rm el-b} = 1.12$
(1.23) for $\omega_{\rm ln} = 40$, respectively. Both sets of values are again
in an intermediate coupling regime.  Based on this, a $T_\mathrm{c}= 52$~K
might be readily achieved e.g.\ by adding a weak coupling to residual interband
spin fluctuations, or with crystal field excitations (CFE) specific for rare
earth systems. The latter can be either attractive or repulsive
\cite{Fulde1978}.  Note that such a weak coupling with conduction electrons at
3~meV has been observed in the antiferromagnetic superconductor HoNi$_2$B$_2$C
in point contact measurements \cite{Naidyuk2007}.  In undoped NdFeAsO,
corresponding peaks at 7.2~meV and 8.6~meV have been observed \cite{Xiao2013}
by inelastic neutron scattering.  In LaFeAsO$_{1-x}$H$_x$, both CFE cases might
be helpful to enhance $T_\mathrm{c}$, depending on the symmetry of the SCOP and
the nature of the CFE. A more sophisticated theoretical study to unravel their
dominant intra- or interband character, as well as low-energy inelastic neutron
measurements, are desirable to settle this issue. Such studies would also be
relevant for the various rare earth-1111 FeSC with the highest bulk
$T_\mathrm{c}$ values apart from FeSe ultrathin films on SrTiO$_3$ substrates
\cite{WangFeSeSTO}, where a high-energy oxygen derived optical phonons of the
substrate may play an important role
\cite{Lee2014,Wang2016,Kyung2016,Choi2016,Kulic2016}.

There are also general arguments against any super-strong coupling regime
formally allowed within Eliashberg-theory: namely, an instability of the
corresponding para-phase against lattice or magnetic polarons or other
instabilities, which are beyond standard Eliashberg theory. In the vicinity of
structural transitions, anharmonicity might also be significant and can result
in temperature dependent spectral densities and anomalous isotope coefficients
\cite{Inada1992,Freericks1996,Chang2009}.  Notice that the highest coupling
observed to date in a confirmed phonon-mediated superconductor ($\lambda_{\rm
el-ph} \leq 2.9$) occurs in amorphous PbBi-based superconductors
\cite{Bergmann1973,Rainer1973}.  A lattice instability might be thwarted there
by additional barriers stabilizing the metastable glassy state.  In this
general context, the recent observation of strong Fe-As bond fluctuations in a
double-well potential by EXAFS measurements on undoped and superconducting
LaFe$_{1-x}$Co$_x$AsO single crystals \cite{Ivanov2016} is of interest. There,
$T$-dependent tunneling frequencies at about 24.6 and 26.7 meV (6 and 6.5 THz)
were reported at low-temperature.  Linking these observations to novel
intraband couplings or polaronic effects in the H-doped La-1111 systems will
require more sophisticated theoretical tools and comparable experimental data.

To summarize this section, we believe that no FeSCs realizes a low-energy
bosonic coupling regime with $\lambda_{\rm el-b}^{\rm low} > 1.5$, with the
possible exception of the rare earth 1111 high-$T_\mathrm{c}$ ($\sim 50$~K)
compounds.  To the best of our knowledge, no high-quality specific heat data
are available for this system due to the small size of the available single
crystals.  But even in this case, an intermediate coupling scenario cannot be
excluded at this time. 

\section{AFe$_2$As$_2$, A= K, Rb, Cs  puzzles - a challenge for theory.}
The AFe$_2$As$_2$ families with A = K, Rb, Cs form a special group among the
FeSCs.  They all have rather low $T_\mathrm{c}$ values of only few Kelvin,
which decreases in moving from K- to Cs-122 \cite{Eilers2016}.  The small and
decreasing $T_\mathrm{c}$ is attributed to an increasing proximity to a QCP,
where fluctuations produce a mass enhancement but no compensation by the
pairing interaction. 

The phase related to the QCP has not been identified yet, but it is unlikely to
be related to a Mott phase. K-122, Rb-122, and Cs-122 are the most heavily
h-doped Fe-pnictides, formally achieving a Fe$^{+2.5}$ valence state, midway
between Fe$^{+2}$ and Fe$^{+3}$.  The former occurs in the magnetic parent
compounds Ba(Sr,Ca)-122, whereas the latter is expected to form an
antiferromagnetic Mott-insulator. Generally speaking, the Fe$^{2+}$ valance
state should be less correlated than the Fe$^{3+}$, but both should realize
{\it commensurate} magnetic phases.  Instead, 
one observes soft {\it incommensurate} magnetic fluctuations 
in K-122 at $T=12$~K, where it appears as a broad peak in the spin
susceptibility centered at $\sim 8$~meV \cite{Lee2011}.  (This energy scale is
significantly lower than the resonance mode energy employed in most of the
strong coupling simulations mentioned above \cite{Ummarino2009,Popovich2010}.)
BaMn$_2$As$_2$ and LaMnAsO have five electrons in their 3$d$-shells
\cite{Zingl2016} and are therefore expected to be in the vicinity of a
Mott-insulator phase. These systems have very high Ne\'el temperatures of
$T_{\rm N}=625$~K and 350~K, respectively.  This is in sharp contrast with the
three compounds discussed here, so it is unlikely that these materials are in
the vicinity of a Mott phase.  Therefore another magnetic phase should be
considered as a candidate responsible for the QCP.

We would like to stress that the central theoretical problem in our opinion is
not whether there are strongly correlated electrons (likely the 3$d_{xy}$
states) present in these compounds. The Curie-Weiss tails of the measured
magnetic susceptibilities at high temperatures $T>T^*$ already evidences this
fact \cite{Wu2016}.  Instead, more subtle questions must be addressed, namely
which of the subsystems is responsible for the QCP, which determine the nature
of the magnetic phase beyond, and which form the bands where superconductivity
occurs? And, how strong are the correlation effects in the corresponding
subsystems? Thus K-, Rb-, and Cs-122, confront us with a complex multicomponent
problem that has not been examined in detail from a theoretical point of view.  

To provide a clear starting point for a discussion of related aspects
given below, we performed various high-precision density functional theory
(DFT) calculations for the densities of states (DOS), as well as the Fermi
surface sheets including a wave vector and orbital-resolved analysis of the
mass anisotropies. Our relativistic DFT electronic structure calculations
were performed using the full-potential FPLO code \cite{fplo,fplo2}.  For the
exchange-correlation potential, both the local density approximation (LDA)
within the parametrization of Perdew-Wang \cite{Perdew1992} or the general
gradient approximation (GGA) with the parametrization of Perdew-Burke-Ernzerhof
\cite{PBE} have been chosen.  The spin-orbit (SO) coupling was treated
non-perturbatively by solving the four component Kohn-Sham-Dirac equation
\cite{KSD}.  The final calculations were carried out on a well-converged mesh
of maximal $10^6$ $k$-points ($100\times100\times100$ mesh, but at least a
$72\times72\times72$ mesh was used) to obtain correct band structure and Fermi
surface information. For all calculations we used the experimental crystal
structures. 

\begin{figure}[t]
\centerline{\includegraphics[width=0.8\columnwidth]{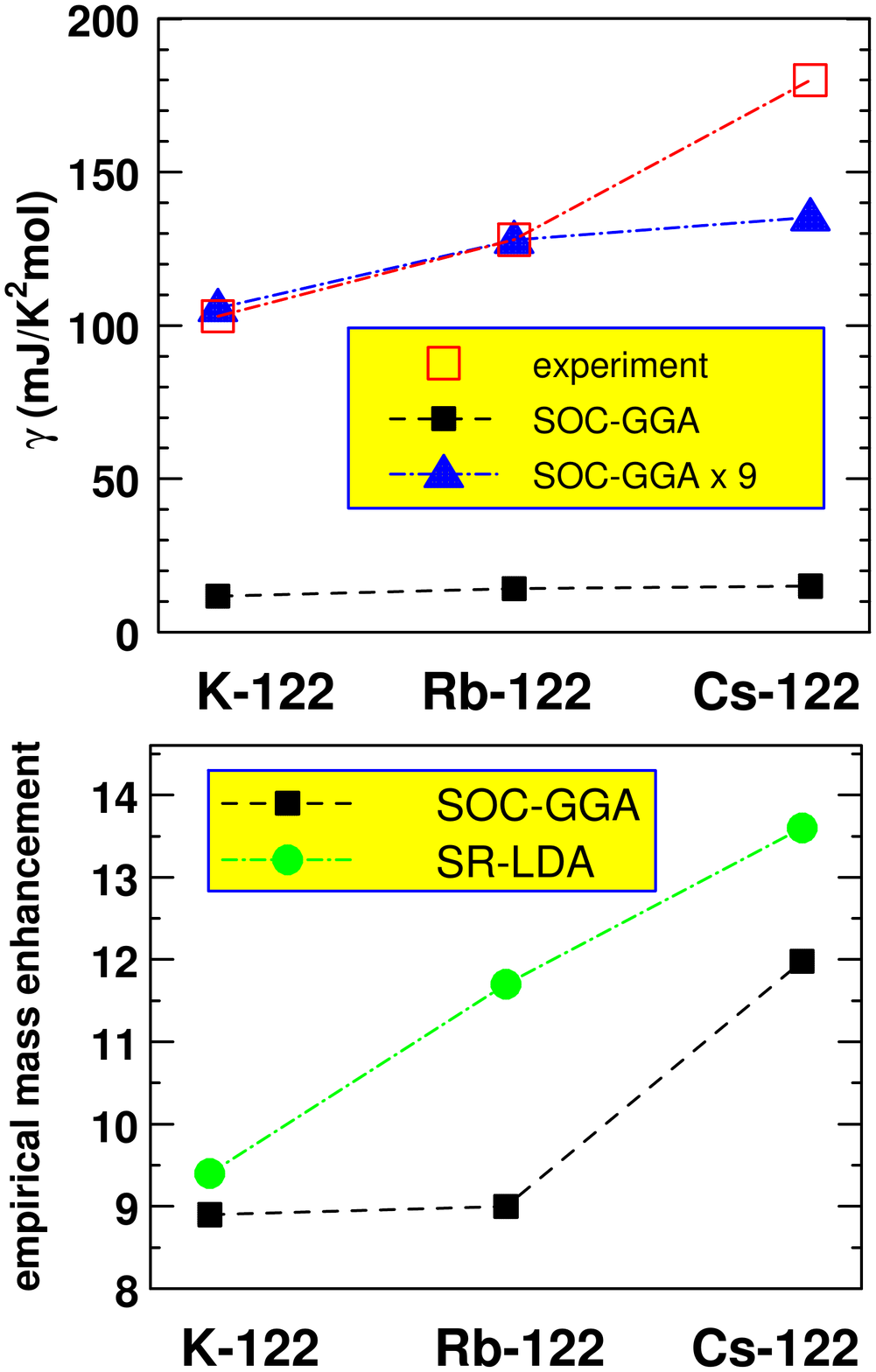}}
\vskip -0.75cm
\caption{({color online})
(a) Bare Sommerfeld constant in (mJ/K$^2$ mole)
according to our GGA calculation with SOC fully  included as compared with
experimental data for KFe$_2$As$_2$ (K), RbFe$_2$As$_2$ (Rb). 
and CsFe$_2$As$_2$ (Cs). (b) The same for the empirical mass enhancement
obtained using scalar relativistic (SR) 
LDA and SOC-GGA calculations for total density of 
states (DOS) $N(0)$ at the Fermi level
as reference quantity. The lines are guides for the eyes.
For the experimental $\gamma$-values please see e.g.\ Refs.\
\cite{Zhang2015,Hardy2016,Wang2016,Khim2016} and the main text.
}
\label{Fig:Sommerfeld}
\end{figure}

The K-, Rb-, and Cs-122 systems exhibit the largest Sommerfeld coefficients
$\gamma$ among the FeSCs, being comparable with of those found for
heavy-fermion systems (see Fig.\ \ref{Fig:Sommerfeld}a).  This observation,
together with a claimed universal Knight-shift scaling anomaly in the NMR
spin-lattice relaxation rate $1/T_1 \propto T^{0.75}$ at $T <T^*$ \cite{Wu2016}
(with $T^*$= 165, 125, and 85~K for K-, Rb-, and Cs-122, respectively), has
lead to proposals for a related emergent Kondo-lattice like scenario for the
whole series \cite{Wu2016}.  But if one uses the DFT calculated DOS
$N(0)=5.0$~states/eV/f.u. (taking into account the SOC)  and a measured
Sommerfeld coefficient $\gamma \approx 103$~mJ/K$^2\cdot{\rm mol}$ for
KFe$_2$As$_2$, one arrives at a large total mass renormalization of $1+\lambda
=8.9$. This value is  very close to that of RbFe$_2$As$_2$: 8.8 (9.0), which is
obtained using the experimental values of $\gamma = 125 (128)$~mJ/K$^2\cdot{\rm
mol}$ and the calculated $N(0)=6.04$~states/eV/f.u..  In other words, these
materials have almost equal mass renormalizations within the experimental error
bars.  This contradicts the notion that the mass enhancement increases
continuously along the series from K- to Cs-122 as reported in Refs.\
\cite{Hardy2016,Eilers2016} (see Fig.\ \ref{Fig:Sommerfeld}b). 

Our SOC GGA calculations yield the closest distance between the Fe 3$d_{xz}$-
and 3$d_{yz}$-derived Van Hove singularities (VHS) and the Fermi level, as well
as the largest DOS $N(0)$ at $E_\mathrm{F}$ as compared to LDA. Therefore, our
analysis will focus on the SOC-GGA calculations to obtain conservative
estimates for the corresponding renormalizations. Then, only CsFe$_2$As$_2$
with $\gamma \approx 180$ mJ/K$^2\cdot{\rm mol}$ and
$N(0)=6.\textcolor{green}{4}$~states/eV/f.u.  shows a surprisingly increased
mass renormalization $1+\lambda = 11.98$, which is less dramatic than suggested
in Refs.\ \cite{Hardy2016,Eilers2016,Mizukami2016}.  (In  Ref.\
\cite{Mizukami2016} for example, a mass renormalization  of about 13 was
reported for Cs-122 using the GGA-PBE within the WIEN2k package.) Note that in
both of our GGA-calculations (scalar relativistic and with SOC included) we
observe an almost twice as large increase of the bare DOS (28 \%) in going from
K-122 to Cs-122 as compared with 15\% reported in Ref.\ \cite{Mizukami2016} and
21\% relative to Rb-122.  We ascribed these discrepancies to our use of an
unusually dense mesh of ${\bf k}$-points ($N_{\bf k}=72\times72\times72$) in
the irreducible BZ.  Note, that both LDA calculations produce a significantly
smaller value of $N(0)$ (see Fig.\ \ref{Fig:DOS_Cs}).  In fact, using our LDA
codes, we would  arrive at a somewhat larger and smoother increase in the mass
renormalization in  going from K-122, Rb-122, and Cs-122: $1+\lambda = 9.4$,
11.7 and 13.6, respectively (Fig.\ 3b).

An inspection of our calculated in-plane and out-of-plane plasma frequencies
shows that the electronic structure according to the GGA-codes is slightly less
anisotropic than those obtained within the LDA-codes.  Within our approach, the
SOC increases the hybridization of the electrons at the $\varepsilon$-Fermi
surface with that on the stronger correlated $\beta$-Fermi surface. Then the
SOC might also contribute to the position of the VHS, and the SOC-GGA provides
the closest distance to $E_\mathrm{F}$. Thus, a weaker ``high-energy" el-el
interaction is needed to explain the closer observed positions of the VHS (on
the order of 10~meV, only). We will discuss this issue elsewhere in the context
of new ARPES data. 

\begin{figure}[b]
\centerline{\includegraphics[width=0.99\columnwidth]{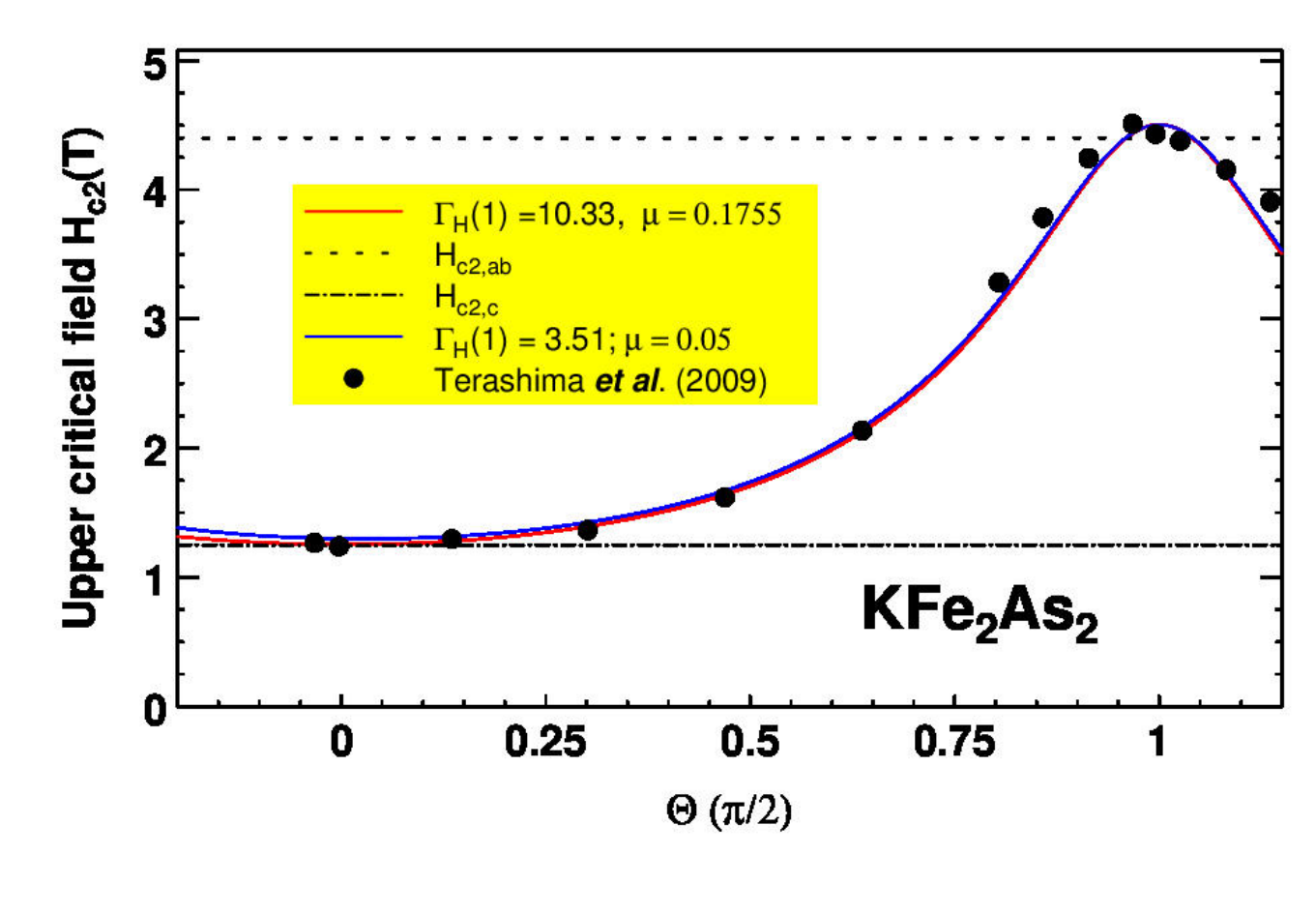}}
\vskip -0.5cm
\caption{({color online}) The anisotropy of the uppper critical field
$H_{c2}(\Theta)$ at low-$T$ taken from Ref.\ \cite{Terashima2009}, where
$\Theta$ denotes the  tilting angle of the magnetic
field relative to the $c$-axis. The red line is the fit using our
semi-analytic theory with an orbital mass anisotropy suggested
by the calculated mass anisotropy of the $\varepsilon$ (blade) Fermi surface sheet
(see Fig.\ 14)
and $\mu$ is the fitted pair-breaking parameter using
the experimental $H_{c2}(T)$ data for H $\parallel$ to the  $ab$-plane
by F.\ Eilers
\cite{Eilers2014} at intermediate
and low-$T$ except a small region below $T_\mathrm{c}$
where the four weakly coupled Fermi surfaces still participate in the superconductivity. }
 \label{terashima}
\end{figure}

In addition to the total mass renormalization, there are significant unexpected
differences between these three compounds, even when they have similar doping
levels and lattice structures.  With respect to superconductivity, the symmetry
of the nodal order parameter ($d$-wave vs.\ $s$-wave with accidental nodes) is
still under debate
\cite{Hardy2013,Hardy2016,Reid2012,Vafek2016,Grinenko2014a,Abdel-Hafiez2013,Okazaki2012,Ota2014}.
It is still unclear which of the five Fermi surfaces sheets plays a dominant
role in the superconductivity and which of them are of minor relevance. Our DFT
calculations of the in- and out-of-plane partial plasma frequencies find that
each of the Fermi surface sheets has a distinct mass anisotropy. This result is
consistent with an analysis of the $T$-dependence of the upper critical field
anisotropy $\Gamma_H=H^{ab}_{c2}/H^{c}_{c2}$.  Taking into account the ideal
cos$^2\Theta$ angular dependence at low-temperature \cite{Terashima2009} (see
Fig.\ \ref{terashima}) \be
H_{c2}(\Theta ,\mu) = \frac{\Gamma_H H^{orb}_{c2,ab}}
{\sqrt{\Gamma_H^2\left(\cos^2\Theta + 2.4\mu^2(T)\right) +\sin ^2 \Theta }},
\ee\label{Eq:Field}
one finds an unexpected simple angular dependence for a multiband
superconductor with rather different mass anisotropies of each of its Fermi
surface sheets.  For the sake of simplicity a  vanishing pair-breaking for
$\Theta=0$ has been adopted in our analysis by Eq.\ (6), in accord with the
experimental data \cite{Terashima2009,Eilers2014,Khim2016} within the
errorbars.  Here, $\Theta$ measures the tilting of the external field direction
from the $c$-axis and $H^{orb}_{c2,ab}$ is the orbital limited upper critical
field without paramagnetic effects and $\mu (T)$ denotes the pair-breaking
parameter (notice that $\mu(0)=\alpha_M/1.76$, where $\alpha_M$ is the
frequently used Maki parameter) scaled by the $d$-wave gap function $\Delta(T)$
and the orbital limited (WHH) upper critical field \cite{Amano2015}
\be
\mu(T)=\mu(0)\frac{\Delta(T)}{\Delta(0)} \frac{H^{orb}_{c2}(T)}{H^{orb}_{c2}(0)} \ .
\ee

At variance with the numerical investigation of Sr$_2$RuO$_4$ performed in
Ref.\ \cite{Amano2015} (which is also a multiband superconductor with a single
dominant band), we have used high-quality analytical approximations for the
latter two functions.  From the observation shown in Fig.\ \ref{terashima}, we
may conclude that only a single Fermi surface sheet survives at low $T$ and
high magnetic fields.  We may identify this sheet with the blade-
(propeller)-like Fermi surface (usually denoted as $\varepsilon$-Fermi surface
sheet in dHvA-measurements) near the corner of the Brillouin zone (see Fig.\
\ref{Fig:FSS_K122}). Thus, we suggest that this sheet of the Fermi surface
is the main locus of superconductivity in K(Rb,Cs)-122, in accord with Refs.\
\cite{Hardy2013,Hardy2016}. But in sharp contrast, this band exhibits a nodal
$d_{x^2-y^2}$-superconducting gap in our
opinion \cite{Grinenko2014,Efremov2016}. In the minor remaining bands in an
ambient field, however, a weak superconducting order parameter is induced which
is, sensitive to competing {\it nonsuperconducting} order parameters. As a
result, the inner Fermi surface sheet might be fully gapped in the case of
coexistence.  Or, in the case of non-coexistence realized on the second Fermi
surface sheet, it occurs in the four nodal regions of the $d$-wave
superconducting order parameter, leading to an octet nodal structure in
accord with high-precision laser ARPES measurements
\cite{Okazaki2012,Ota2014}.  Unfortunately, these same experiments cannot probe
the $\varepsilon$-Fermi surface sheet and hence its gap structure. Further
experimental and theoretical studies are necessary to settle this subtle
puzzle. 

 The quenching of the minor bands is probably related to the non-BCS shapes of
$\Delta(T)$ for these bands, which is generic for weak interband coupling. Then
the significantly suppressed gap amplitudes near $T_\mathrm{c}$ and the
enhanced magnetic susceptibility of the most correlated $\beta$-band may
dramatically enhance the paramagnetic effect provided by the Zeeman-splitting
\cite{Fuchs2009}. This view would explain why the most anisotropic band
quenches first and $\Gamma_H$ is reduced already in the very vicinity of
$T_\mathrm{c}$. The calculated moderate mass anisotropy of the blade
$\varepsilon$-band is in accord with the significant $k_z$ dependence of the
extremal cross sections observed in de Haas-van Alphen (dHvA) measurements
reported by Yoshida {\it et al.} \cite{Yoshida2014} (1.29 \% including the
Z-point and 0.89 \% including the $\Gamma$-point of the BZ in units of its
total area).  The significant deviation of the absolute values in the DFT
calculations by a factor of 10 reflect the band shifts discussed below in the
context of the positions of van Hove singularities. 
\vskip 0.5cm 

\subsection{An unusual low-energy bosonic excitations in K-122.}
Recent work involving some of the current authors observed a clear bosonic peak
near 20~meV in several ballistic point-contact measurements of K-122
\cite{Naidyuk2015}.  The standard interpretation for this peak as being due to
spin fluctuations or harmonic phonons can be ruled out. Our calculation of the
harmonic el-ph Eliashberg function $\alpha^2F(\nu)$ revealed no sharp peak
structures at this energy. Regarding spin fluctuations, the available inelastic
neutron scattering data \cite{Lee2011} shows a broad peak structure in the spin
susceptibility, but at a lower energy $\sim 8$~meV. We are therefore forced to
consider other candidates

\begin{figure}[t]
\centerline{\includegraphics[width=1.0\columnwidth]{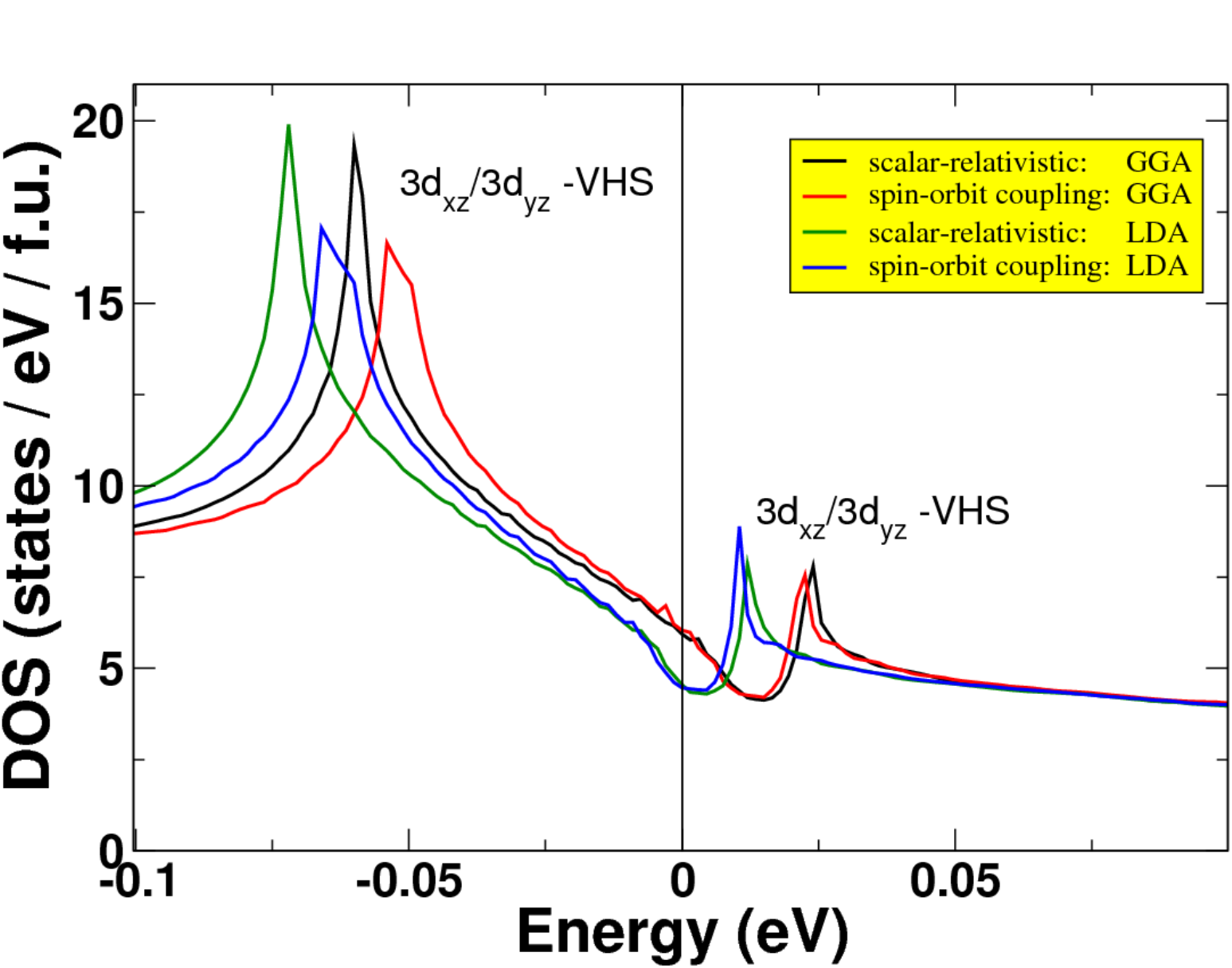}}
\centerline{
\includegraphics[width=1.0\columnwidth]{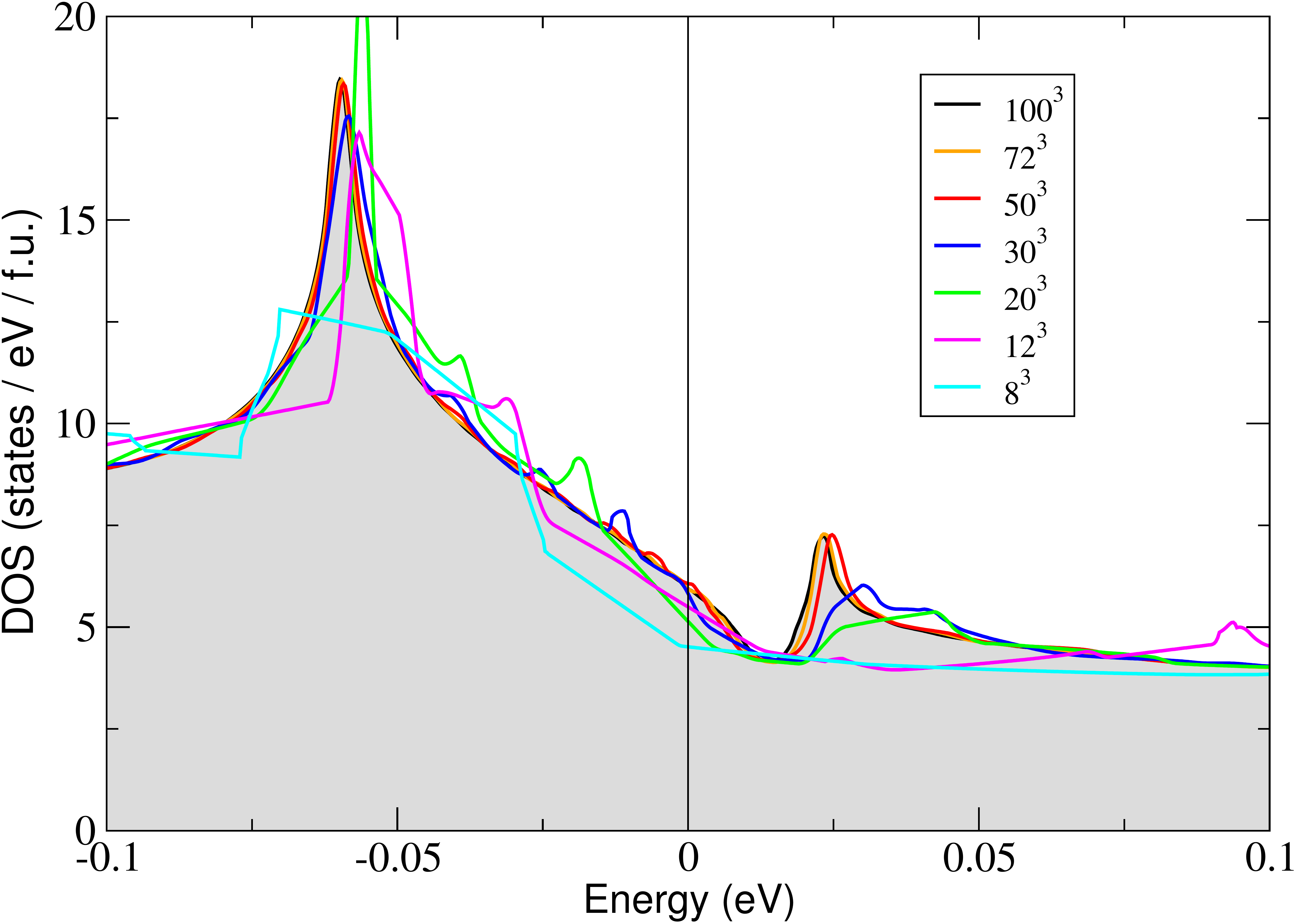}
}
\vskip -0.1cm
\caption{({color online}) Upper panel: Calculated total density of states (DOS) for
RbFe$_2$As$_2$ within scalar relativistic and spin-orbit coupling included
 LDA and GGA codes, respectively, using a dense mesh of
${\bf k}$-points (72x72x72).
Notice the pronounced peak--like feature
at the bottom of the empty el-pocket near +15~meV, which stems from a generic quasi-2D VHS. Lower panel: Dependence on the number of ${\bf k}$-points for the scalar relativistic GGA calculation.}
\label{Fig:DOS_Rb}
\end{figure}

\begin{figure}[b]
\centerline{\includegraphics[width=1.0\columnwidth]{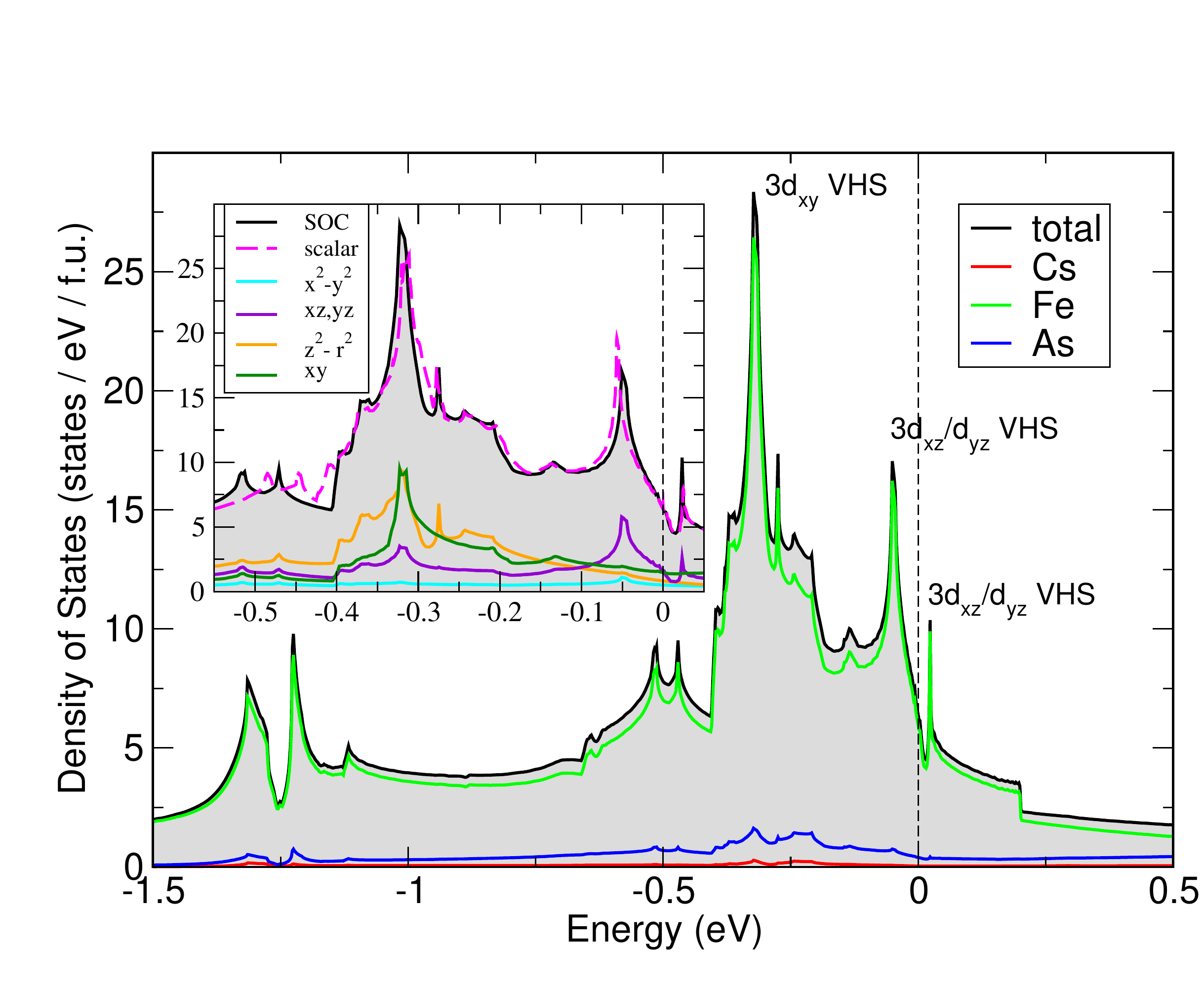}}
\vskip -0.3cm
\caption{({Color online}) Exemplarily calculated total density of states (DOS) for
CsFe$_2$As$_2$ within full relativistic (SOC) DFT-codes.
Inset: orbital-resolved DOS.
Notice also the weak VHS near the
bottom of the unoccupied el-pocket at about 20~meV discussed in the Sect.\
5.1. and mentioned also in the caption of Fig.\ \ref{Fig:DOS_Rb}. }
\label{Fig:DOS_Cs}
\end{figure}

One possibility initially advanced in Ref.\  \cite{Naidyuk2015}, is that the
20~meV boson is a related to low-lying exciton, involving an electron transfer
from one nested $\Gamma$-centered holelike band to an empty electronlike band
located slightly above the Fermi energy and centered near the BZ corner. This
electron band is generic to all FeSCs and it becomes completely unoccupied with
hole-doping at a corresponding Lifshitz-transition, which occurs at $x \approx
0.7$ -- $0.8$ in Ba(Sr)$_{1-x}$K$_x$Fe$_2$As$_2$.  Our DFT calculations place
this band about 10 to 25~meV above the Fermi level and further show that its
band minimum produces a weak van Hove singularity (VHS) in the PDOS, as shown
in Fig.~\ref{Fig:DOS_Rb}.  Here, calculations must be carried out on
sufficiently dense ${\bf k}$-grids to resolve it, e.g.  $N_{\bf
k}=72\times72\times72 = 72^3$ ${\bf k}$-points in the irreducible part of the
BZ.   This is illustrated in Fig.\ \ref{Fig:DOS_Rb}, where we show DOS results
for K-122 as a function of $N_{\bf k}$.  In the present case, at least $N_{\bf
k} = 50^3$ are needed to achieve the necessary convergence.  Usually, we have
used $N_{\bf k} =72^3$ and checked that $N_k > 100^3$  gives practically the
same result, as shown in Fig. \ref{Fig:DOS_Rb}.   In addition to the visible
impact on the sharpness of the VHS, we find that the ${\bf k}$-grid density
also affects the value of the DOS at the Fermi level and deviations of up to 20
\% of the converged value can be obtained for less dense grids. To the best of
our knowledge, this significant ${\bf k}$-grid dependence has not been taken
into account explicitly when estimating the strength of many-body effects in
the pnictide literature although it is well-known in principle.  Like the
${\bf k}$-mesh, the details of the lattice structure such as the As position
have a significant impact on the physical properties. Generally, low-$T$
structural data are used (if available) to analyze e.g.\ the Sommerfeld
constant $\gamma_\mathrm{b}$ correctly. If one uses room temperature data with
enlarged lattice constants to compute $\gamma_\mathrm{b}$, which is then
compared to low-$T$ measurements, then the correlation effects can be
overestimated.  In the case of A-122 (A = K, Rb, and Cs), low-$T$ structural
data (down to $T=1.7$~K) are only available for K-122 \cite{Avci2012}.  To
circumvent this problem for Rb-122 and Cs-122, we scaled the available room
temperature data for the latter \cite{Eilers2014} in a similar way as in K-122.

\begin{figure}[t]
\centerline{\includegraphics[width=0.7\columnwidth]{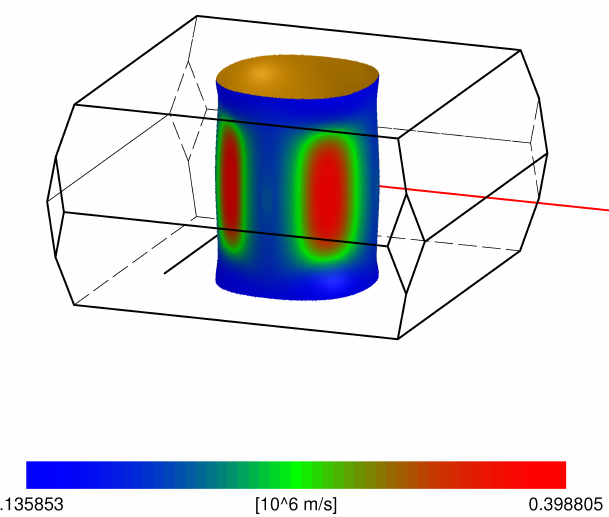}}
\centerline{\includegraphics[width=0.7\columnwidth]{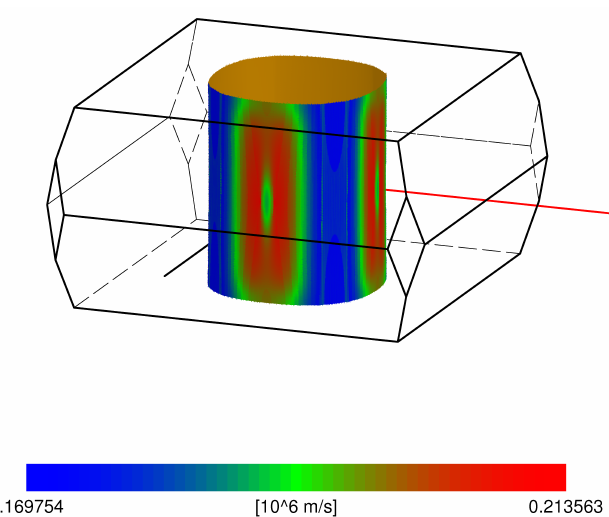}}
\centerline{\includegraphics[width=0.8\columnwidth]{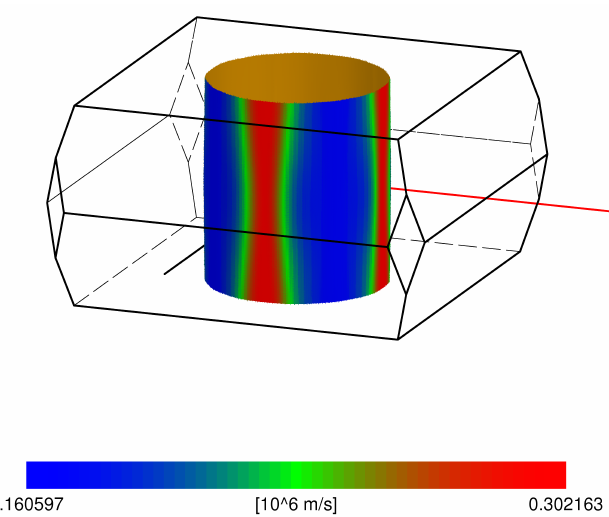}}
\caption{(Color)
The three FSS of K-122 centered around the $\Gamma$-point of the Brillouin zone (BZ)
within DFT.
Upper: FSS with dominant 3$d_{z^2}$-character. Middle: FSS with
dominant 3$d_{xy}$-chatcter.
Lower: FSS with dominant 3$d_{xz}/3d_{yz}$-character
The indicated colors measure the values of the unrenormalized
Fermi-velocities.}
\label{Fig:FSS}
\end{figure}
\begin{figure}[b]
\centerline{\includegraphics[width=0.7\columnwidth]{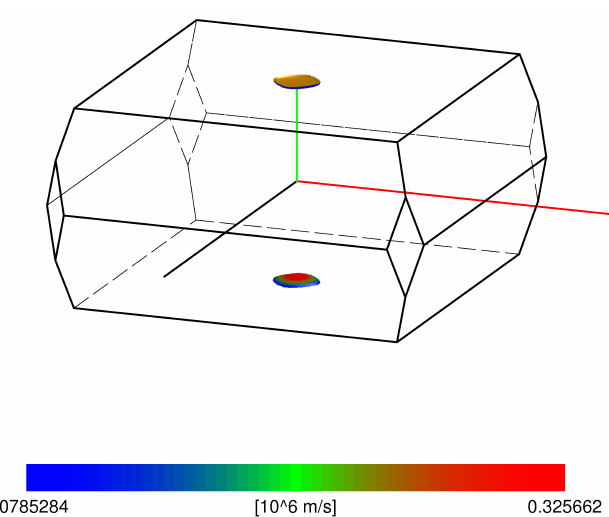}}
\centerline{\includegraphics[width=0.7\columnwidth]{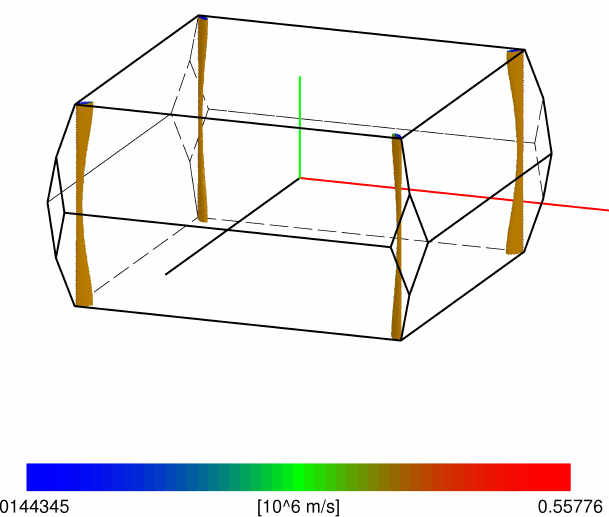}}
\centerline{\includegraphics[width=0.7\columnwidth]{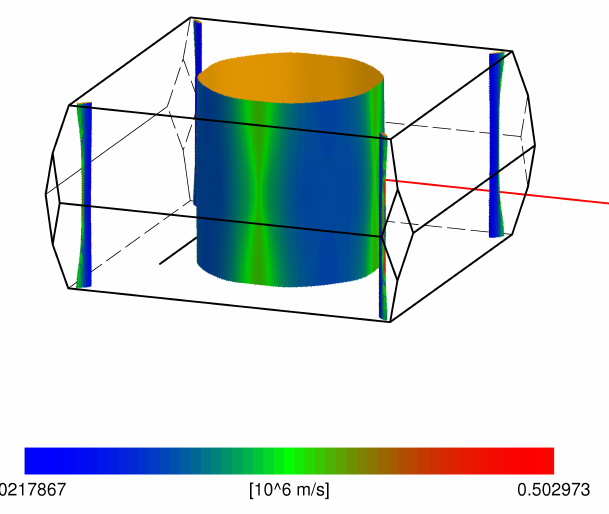}}
\caption{(Color) 
Upper panel: the closed FSS of K-122 around the Z-point of the BZ with dominant
3$d_{z^2}$ character.} Middle:  
The blade-like FSS near the corners of the BZ for K122 according to DFT.
Notice the smaller (larger) cross section  at $k_zc$ =0 ($\pi$), respectively, corresponding
to planes which contain the $\Gamma$ and the Z-points of the BZ.
Lower panel: The same for Cs-122 where another highly anisotropic
FSS with 3$d_{xz}, d_{yz}$ orbital character centered around the 
$\Gamma$-point of the BZ corresponding to the lower panel in Fig.\ 14 is shown, too. 
The indicated colors measure the values of the unrenormalized
Fermi-velocities.
\label{Fig:FSS_K122}
\end{figure}

We suggest that this VHS could affect some physical properties in the vicinity
of the Lifshitz transition at $x\sim0.7$ -- $0.8$.  The observation of an
exciton is somewhat unusual for an ordinary metal since the Coulomb interaction
between the excited electron, and the remaining hole is usually entirely
screened. As a result, the exciton becomes ill-defined as a quasi-bosonic
excitation interacting with the other conduction electrons.  The heavy masses
for these three systems under consideration here, however, may be helpful for
stabilization of such excitons through a reduction of dynamical screening.  

Alternatively, the 20~meV boson may be related to other types of excitations.
In the context of charge excitations, two recent studies devoted to K-122
\cite{Wang2016a} and Rb-122 \cite{Civardi2016} are of interest. In Ref.\
\cite{Wang2016a} critical spin {\it and} charge fluctuations have been reported
using pressure measurements.  In Ref.\ \cite{Civardi2016} a partial charge
order below 20~K was ascribed to an electronically driven phase separation,
which is accompanied by a small gap of 17~K in the spin-lattice relaxation rate
$1/T_1$ probed by NQR measurements.  Alternatively, the possibility of orbital
excitations or low-lying plasmon excitations (due to the presence of both very
heavy and lighter electrons) should be investigated theoretically. 

Another option is the vicinity of a QCP  of a novel SDW phase with magnetic
stripes (discussed in the next section) or strong anharmonicity of the Fe-As
bond phonons \cite{Ivanov2016}. The latter has been observed in electron
overdoped BaFe$_{2-x}$Co$_x$As$_2$ system with $x=0.11$, above the ordinary
spin-stripe phase.  The tunneling mode of about 5~THz ($\approx 20.4$~meV)
derived there is fascinating since it nearly coincides with the unknown bosonic
mode's energy and suggests the involvement of a strongly anharmonic phonon
mode.  To confirm such a speculation, it would be desirable to have (i)
corresponding EXAFS measurements on a K-122 single crystal and (ii) a deeper
theoretical understanding of the suggested double-well potential \cite{Berciu}.
In such a scenario, a hybridization with the above discussed excitonic mode or
another charge excitation might also occur. Thus inelastic neutron scattering
and Raman measurements, as well as isotopic substitutions, might provide a
clearer picture of the nature of this bosonic mode.   {\em Ab initio} DFT
calculations are hampered here by the fact that they cannot reproduce the
experimental pnictogen position, which is known to affect many physical
properties of the FeSCs \cite{Kuroki2010,Mizuguchi2010}.  The previously
mentioned unusual dynamical properties of the Fe-As bond provides additional
interest to the microscopic consequences for superconductivity and competing
phases. 

Given all of this, the elucidation of the fate of a corresponding bosonic
excitation in Rb- or Cs-122 is a challenge for both experiment and theory. 

\subsection{Orbital Selective Mott physics versus shifted van Hove singularities.}\label{Sec:Shifted} 
STM measurements \cite{Fang2015} on K-122 have observed a VHS related to the
3$d_{xz}/3d_{yz}$ orbitals. The observed peak exhibits a rather asymmetric
shape resembling our DFT calculations shown in Figs.\ \ref{Fig:DOS_Rb} and
\ref{Fig:DOS_Cs} for Rb-122 and Cs-122, respectively, but squeezed in width by
a factor of two.  (Due to a much smaller number of ${\bf k}$-points, some of
these VHS and others remained unresolved in Refs.\
\cite{Eilers2016,Mizukami2016}.) Moreover, a large discrepancy between
experiment and DFT predictions (by a factor of five) appears when comparing its
position about the Fermi level, where the experimental features are found
closer towards $E_\mathrm{F}$. This trend is in agreement with the predictions
of DMFT studies, where the Hubbard $U$ and the Hund's coupling $J$ are more
accurately taken into account.  In this context, and that of FeSe, an intersite
Coulomb interaction $V$ (usually ignored in the DMFT) may also contribute to
significant upshifts of the VHS from -250 meV within the DFT to about 25 meV
for FeSe, as observed in ARPES measurements.  This way the Hubbard $U$ and
Hund's exchange $J$ might be somewhat reduced as compared to the values adopted
so far.  SOC can be important for determining the position of the VHS and the
value of the DOS; the two-Fe unit cell accounts for the two different As
positions and thus provides an additional SOC between the 3$d_{xy}$ and the
3$d_{xz}/3d_{yz}$ orbitals at nearest neighbor Fe sites \cite{Fernandes2016}. 

We will discuss the analogous situation in K-122 (where these effects are more
pronounced) and its sister compounds in greater detail elsewhere, along with
the inclusion of new ARPES data and DMFT calculations.  As a consequence, the
filling ratios of the bands containing the most strongly correlated orbitals
are reduced.  For completeness, we note that for Cs-122, a VHS close to
$E_\mathrm{F}$ has not been resolved \cite{Yang2016} so far. Instead a turning
point in resistivity and specific heat data at $T^*=13$ ~K, an anomalous $T^3$
electronic specific heat below $T^*$, (pseudo) gap-like features at -3.7~meV
and 4.7~meV well above $T_\mathrm{c}=2.11$~K, and an unusually large
superconducting gap value of 1.2~meV below $T_\mathrm{c}$ (pointing possibly to
pronounced strong coupling and quantum criticality in the normal state
\cite{Zaanen2009}) have been derived from specific heat and STS measurements,
respectively.  This puzzle of the missing VHS in Cs-122 could be explained by
substantial orbital fluctuations related to a locally broken tetragonal
symmetry for which a splitting into two components one above and a counterpart
below $E_\mathrm{F}$.  Alternatively, it might be related to the opening of a
pseudogap \cite{Wu2016} and or to a charge ordering suggested for an
electronically-driven phase separation like in that suggested for Rb-122
\cite{Civardi2016} at $T=20$~K.  The very observation of charge ordering
provides strong arguments against the vicinity of a Mott phase. As an
experimental example illustrating that point, we refer to the A15 phase of
Cs$_3$C$_{60}$ \cite{Alloul2016}.  The much lower $T_\mathrm{c}$ is attributed
to the nodal $d$-wave character in the present case.  The active presence of an
incipient band in the superconducting state might explain the smaller
anisotropy of $H_{c2}$ as compared to the sister compounds K-122 and Rb-122,
despite the largest spacing between the adjacent FeAs-planes due to the biggest
ionic radii of the Cs$^+$ ions. 

\begin{figure}[b]
\centerline{\includegraphics[width=1.0\columnwidth]{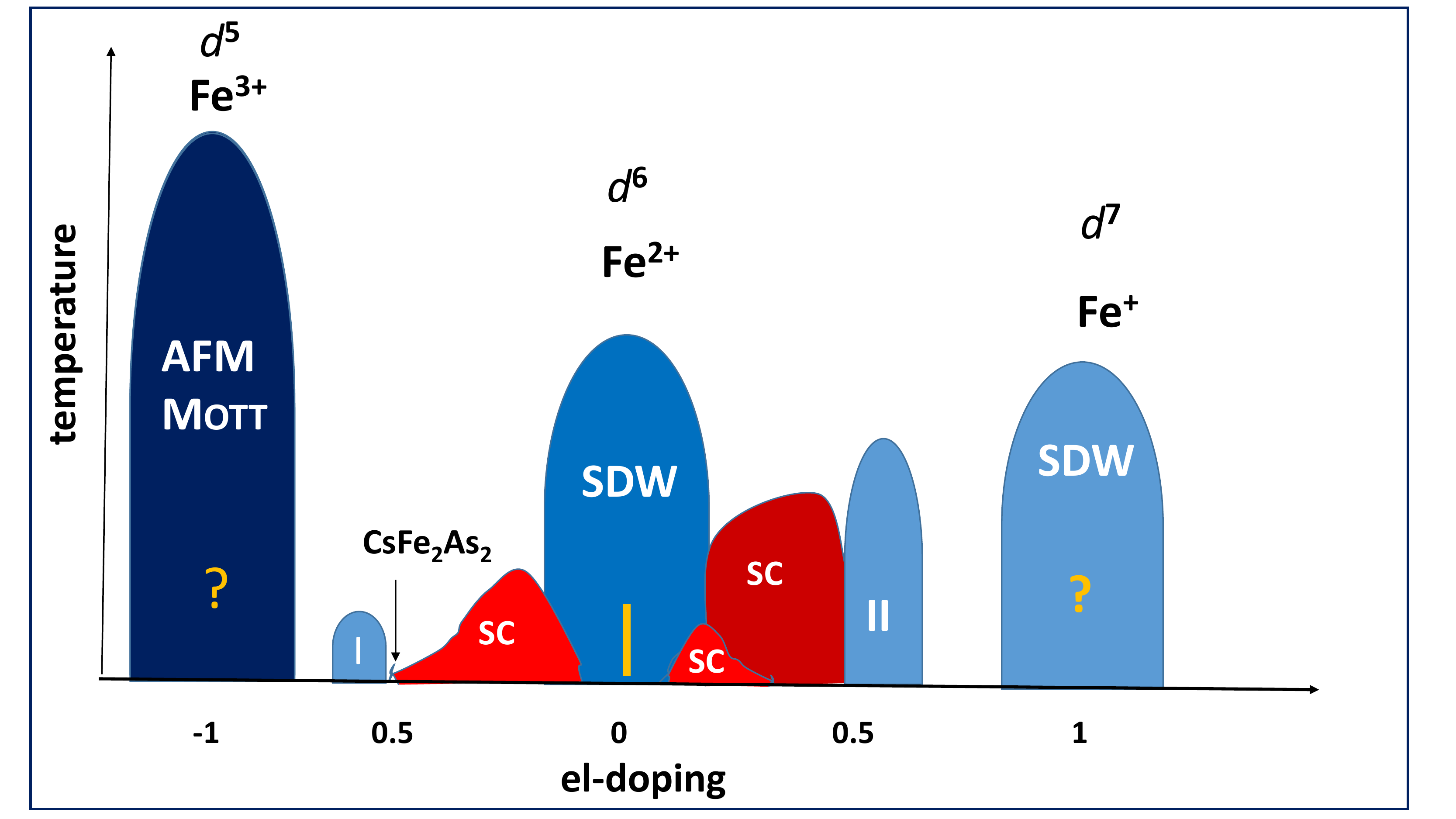}}
\caption{(Color) Suggested general phase diagram of Fe pnictide compounds.
Blue: magnetic regions, red: superconducting regions. Phase I - a combined
charge, orbital, and spin ordered phase responsible for the vicinity of the critical
point as discussed in the text. The yellow line at isovalent or no doping
stands for such systems as Li(Na)FeAs, P-doped Ba(Sr)-122 and bulk FeSe where
the competing magnetic SDW magnetic stripe-phase is absent or strongly suppressed.
Phase II has been observed but not been yet characterized experimentally.
The outermost hypothetical phase around Fe$^+$ is our suggestion.
The bright red and dark red regions stand for 122 and H doped La-1111
(under pressure) \cite{Kawaguchi2016}
FeSC  compounds, respectively.
}
\label{Fig:PDFE}
\end{figure}

 In addition, some admixture of non-Fermi liquid behavior might be visible in
the $T$-dependent resistivity data, which shows a subquadratic exponent of 1.7
below 10~K.  Such an unusual exponent points towards a multiband effect with
different Fermi liquid and/or to non-Fermi liquid subsystems.  The observation
of several distinct characteristic temperatures scales (such as the maximum
for the magnetic susceptibility near 50~K, a new scaling of the spin relaxation
rate $1/T_1$ below about 85~K, and a further slight kink of the resistivity at
about 120~K) also points to a multiband picture that cannot be described using
an effective single band model, where a single freezing temperature separates
an (extremely) bad metallic and a Fermi liquid phase.  We speculate that
all these anomalies might be related to the vicinity of a QCP, as proposed in
Ref.\ \cite{Eilers2016} (ignored in Ref.\  \cite{Hardy2016}) and a novel
incommensurate magnetic phase with a Fermi surface reconstruction. (See also
Fig.\ \ref{Fig:PDFE}, where a general phase diagram for the doping dependence
of FeSC and the related neighboring magnetic phases is suggested.)   Here, the
puzzling unresolved VHS might also be affected by Fermi surface reconstruction
and broadening effects due to strong spin-orbit coupling and somewhat enhanced
correlation effects as compared to K-122.

 These considerations show that the positions of the VHS, if visible, can
provide a unique possibility to measure the strength of the high-energy el-el
interactions.  At the same time, the position of the VHS can also influence
estimates for the high-energy mass renormalization.  For example, a shift
and/or the broadening of the VHS can lead to an increase of the DOS at the
Fermi energy $N(0)$. As such, a weaker el-el interaction would be required to
reproduce the large Sommerfeld coefficient $\gamma \approx$ 100~mJ/${\rm
mol}\cdot$K$^{2}$.  Using full relativistic GGA-FPLO calculations, the
experimental $\gamma \approx$ 125~mJ/${\rm mol}\cdot$K$^{2}$ for Rb-122 can be
understood without adopting a stronger el-el interaction at variance with Ref.\
\cite{Eilers2016}. 

 In order to also explain the record value of $\gamma = 180$~mJ/${\rm
mol}\cdot$K$^{2}$ found in the Cs-122 compound, one is left with two options:
(i) introduce a strongly increased Hubbard $U$, as was done in Ref.\
\cite{Eilers2016} (which is not very plausible for closely related compounds;
alternatively, one should adopt a significant reduction of the screening by
almost localized quasi-particles), or (ii) introduce an increased low-energy
el-el interaction caused by the closer vicinity of a detrimental QCP. This
latter scenario was proposed in Ref.\ \cite{Dong2010,Eilers2016}. Here, an
increase of the $\lambda_{\rm el-b}^{\rm low}$ entering mass renormalization
from about 0.5 for K-122 to 1 for Cs-122 could explain the anomalous magnitude
of the Sommerfeld coefficient $\gamma$. At the same time, this would explain
the decrease of $T_\mathrm{c}$ if no corresponding increase in the
$\lambda_{\rm el-b}^{\rm low}$ entering the anomalous self-energy is
introduced. (These two contributions are often denoted $\lambda_z$ and
$\lambda_\phi$, respectively.)  In this scenario, other effects such as a
change of the bosonic spectra favoring competing pairing symmetries could also
play some role.
 
 The low-energy bosonic scenario could also be helpful in resolving another
puzzle related to the smaller magnetic susceptibility of Cs-122 in comparison
to Rb-122 and K-122.  A small and even a decreasing Wilson ratio
 \begin{equation}
 R_W= \frac{4\pi^2k_{\mbox{\tiny B}}\chi(0)}{3g^2J (J+1)\gamma (0)}
 \end{equation}
 is widely used as a measure of the relative strength of correlation
effects.  Starting from $R_W({\rm K}) \approx 1$, and using the data from Ref.\
\cite{Wu2016} (Fig.\ S4 and Tab.\ S1), one estimates $R_W({\rm Rb})/R_W({\rm
K})\approx $0.727 and $R_W{\rm Cs})/R_W({\rm K})\leq $~0.394,  while  less
correlated FeSC exhibit $W_R$-values in between 2 and 5.  This result is in
obvious conflict with the notion that stronger correlations are present in
Cs-122 and Rb-122 when compared to K-122.  Here, we mention that among a
large number of known (until the year 2000) heavy-fermion superconductors,
only three of them (CeCu$_2$Si$_2$, UPt$_3$, and UBe$_{13}$) exhibit Wilson
ratios $W_R \approx 1$ \cite{Radousky2000}.  To the best of our knowledge,
there is no system with such a very low $W_R \sim 0.4$ as observed in Cs-122.
In general, it is very challenging at present to answer the question about
Kondo-like physics presence in these three compounds.  To gain more insight
into this issue, resistivity measurements well above 300~K with the aim of
detecting a maximum of $\rho(T)$ would be helpful. Spin-orbit coupling and an
explicit account of As 4$p$ orbitals might also be necessary for obtaining a
microscopic estimate of the corresponding Kondo exchange coupling $J_K$ between
the suggested almost localized 3$d_{xy}$ spins and the less correlated 3$d_{xz}
/ 3d_{yz}$ and other electrons.

 In accord with NMR data for K-122 \cite{Wu2016} and especially for NQR data
for Rb-122 \cite{Civardi2016}, the FeSC-related $R_W$ puzzles might be resolved
if low-energy {\it charge} fluctuations were also involved in addition to
spin-fluctuations.  Since the former are expected to compete with significant
on-site correlations on the Fe-sites, we suggest that the latter should mainly
involve the much less correlated As 4$p$ states. As a result, the unknown phase
related to the QCP might be a combined magnetically and weakly charged ordered
phase \cite{Drechsler1997} unlike the purely magnetic Mott-phase, as suggested
in Ref.\ \cite{Eilers2016}.  Also, the smaller magnetic moment of 0.97
$\mu_{\rm \tiny B}$/Fe  (compared to 1.48 for K-122 and 1.45 for Rb-122) is
unexpected within the slaved spin and DMFT scenarios without taking into
account either residual correlations in the As $4p$ states or various intersite
Coulomb interactions.
 
Among the three significant VHS, the Fe 3$d_{xz}$/$d_{yz}$ derived VHS is the
closest to $E_\mathrm{F}$ and is of special interest here.  Its position is
between the $\Gamma$- and the $X$-points at the corner of the BZ (i.e.\ on the
nodal line of a $d_ {x^2-y^2}$-wave order parameter). Under these
circumstances, critical low-energy spin fluctuations will not provide more glue
to SC, but merely enhance the $\lambda_z$ entering the mass enhancement
relative to $\lambda_{\phi}$ entering the anomalous self-energy
\cite{Bulut1996,JohnstonDOS}.  This combination would result in a lower
T$_\mathrm{c}$, in accord with the experiment.  
As compared with our first single and two-band calculations in the weak
coupling limit \cite{Abdel-Hafiez2013,Grinenko2014}, by adopting a small
difference $\lambda_{z}$-$\lambda_{\phi}$ (as suggested by the small $T_c$
in the vicinity of such a quantum critical point) a strong coupling case with
$\lambda_{z} \sim 2$ might be realized by adopting a reduced frequency of the
spin fluctuations in both channels. This way a moderate enhancement of
$\lambda_{\phi} > 1$ can be achieved while conserving the observed low
T$_\mathrm{c}$- value.  Low-$T$ specific heat and INS measurements like for
K-122 \cite{Lee2011} are requested for Cs-122  and Rb-122 to confirm such a
scenario.

Upon further h-doping, the VHS might come closer still to $E_\mathrm{F}$.  This
shift could trigger a new incommensurate magnetic phase, distinct from the
commensurate Mott phase predicted by orbital-selective Mott scenarios
\cite{Linscheid2016}. We note that the situation in (otherwise similar)
Sr$_2$RuO$_4$ is very different: in Sr$_2$RuO$_4$ the VHS of 4$d_{xy}$
electrons in the tetragonal phase is located 20~meV below $E_\mathrm{F}$ and
can be split by external pressure or tension.   One of the split components in
the orthorhombic phase moves towards $E_\mathrm{F}$, thereby strongly enhancing
$T_\mathrm{c}$ \cite{Taniguchi2015}.  We ascribe this opposite effect to the
different symmetries of the SCOP in both systems. Regardless, a similar
fine-tuning of the VHS-position would be of great interest in our case,
providing more insight into the competition of various instabilities.  

Returning to Cs-122, we would like to note that strong nematic
fluctuations, as suggested recently on the basis of NMR and NQR
data\cite{Li2016}, would have a similar local splitting effect as magnetic
stripe phase fluctuations. In this context, it is very likely that the critical
phase I denoted in Fig.\ 9 includes components  from all of the spin,
orbital, and charge degrees of freedom. More sophisticated model calculations
and additional experimental studies might elucidate which of them is the
dominant driving instability. In such a complicated situation,  even weak
disorder effects might be responsible for slightly different observations
reported by various experimental groups. The observation of a disordered glassy
magnetic phase coexisting with an only slightly suppressed superconducting
state in earlier single crystal samples of K-122 but with a reduced total
Sommerfeld coefficient $\gamma$, as well as with a significant residual
counterpart at $T \rightarrow 0$ \cite{Grinenko2013}, is rather remarkable in
this context. (Polycrystalline samples also have reduced Sommerfeld
\cite{FukazawaPreprint}.) The former might correspond to nonsuperconducting
gaps in some of the weakly coupled non-dominant FSS discussed above in the
frame of the octet gap picture, and the latter might be related to the quenched
superconductivity in weak external magnetic fields reported above.  Further
studies would be helpful for checking these suggestions. This general
nontrivial impurity aspect, as well as the experimental determination of the
low-$T$ structures for Rb-122 and Cs-122 including defects, might be very
helpful to find a realistic physical scenario for phase I. We argue, however,
that this phase is not directly related to a (selected) Mott-phase and involves
essentially the less strongly correlated 3$d_{xz}$ and 3$d_{yz}$ orbitals and
the $\varepsilon$ (blade) FSS, in particular.

 In conclusion, with increasing h-doping novel magnetic and superconducting
phases might be found, especially, if a $d_{xy}$-superconducting phase in the
very vicinity to a selective Mott or to an Fe$^{3+}$ state would occur. In the
crossover region a gapless $d_{x^2-y^2}+{\rm i}d_{xy}$ phase would also be a
candidate.  Present day sophisticated many-body approaches describe
qualitatively only some aspects of the normal state properties of FeSC and miss
effects such as non-Fermi-liquid behavior in the vicinity of QCPs. 

\section{Conclusions}
We have provided strong evidence for the presence of low-energy electron-boson
interactions that determine the superconducting and normal state properties of
the FeSC. They can be simulated reasonably well by multi-band Eliashberg-theory
(also including impurity scattering effects, see Refs.\
\cite{Efremov2016,Korshunov2016}).  A phenomenological analysis of normal state
and superconducting properties of the FeSC shows that in the retarded glue
scenario the el-boson coupling constant averaged over all Fermi surfaces
$\lambda_{\rm el-b}$ is limited by about 1.5. In other words, no strong
coupling regime is realized so far.  There is also growing evidence for the
presence of non-negligible intraband couplings, which stabilize $T_\mathrm{c}$
and in some cases.  In some cases, this intraband component can even be
dominant, as in the case of nodal 4$d$-wave symmetry or an $s_{++}$ SCOP
induced by interband impurity scattering. Within such a combined scenario,
involving both interband spin-fluctuation mediated interactions and a (not yet
well understood) intermediate intraband interaction, even the highest values of
$T_\mathrm{c}$ observed in el-doped 1111 systems can be at least qualitatively
explained.

\begin{acknowledgement}
The work presented in this review was made possible
by the financial support provided by the German Research
Foundation (Deutsche Forschungsgemeinschaft(DFG))
through priority program SPP 1458. 
V.G.\ is grateful to the DFG through the 
grant GR4667/1-1 for financial support.
The authors thank
B.\ B\"uchner, R.\ Hackl, C.\ Honerkamp and
D.\ Johrendt for initiating this priority program and
for 
discussions.
Further thanks to J. van den Brink, D.~Efremov, O.~Dolgov, 
S.\ Borisenko, S.\ Backes, J.\ Tomczak, L.\ de~Medici,
R.\ Valenti, S.\ Biermann, G.\ Kotliar, A.\ Chubukov, M.\ Dressel, R.\ Thomale,
J.\ Schmalian, H.\ K{\"u}hne, A.\ Boris, A.\ Charetta, A.\ Yaresko, H.-J.\ Klauss,
G.\ Fuchs, R.\ Schuster, C.\ Geipel, N.M. Plakida, K.\ Kikoin,
F. Hardy, S.\ Khim ,
P.\ Mydosh, I.\ Morozov, and S.\ Aswartham for numerous and helpful
discussions.
\end{acknowledgement}

\bibliographystyle{pss}
\bibliography{spp_report_mass_renormalization10}

\providecommand{\WileyBibTextsc}{}
\let\textsc\WileyBibTextsc
\providecommand{\othercit}{}
\providecommand{\jr}[1]{#1}
\providecommand{\etal}{~et~al.}


\begin{thebibliography}{[100]}

\bibitem{Kamihara2006}
 \textsc{Y.~Kamihara},  \textsc{T.~Watanabe},  \textsc{M.~Hirano},  and
  \textsc{H.~Hosono},
 \jr{J.\ Am.\ Chem.\ Soc.} \textbf{128}, 10012 (2006).


\bibitem{Kamihara2008}
 \textsc{Y.~Kamahara},  \textsc{T.~Watanabe},  \textsc{M.~Hirano},  and
  \textsc{H.~Hosono},
 \jr{J.\ Am.\ Chem.\ Soc.} \textbf{130}, 3296 (2008).


\bibitem{Johnston2010}
 \textsc{D.~Johnston},
 \jr{Adv. Phys.} \textbf{59}, 803 (2010).


\bibitem{Paglione2010}
 \textsc{S.\,J. Paglione} and  \textsc{R.~Greene},
 \jr{Nat. Phys.} \textbf{6}, 645 (2010).


\bibitem{Stewart}
 \textsc{G.\,R. Stewart},
 \jr{Rev. Mod. Phys.} \textbf{83}, 1589 (2011).


\bibitem{Hirschfeld2016}
 \textsc{P.\,J. Hirschfeld},
 \jr{Comptes Rendus Physique} \textbf{17}, 197 (2016).


\bibitem{Mannella2014}
 \textsc{N.~Mannella},
 \jr{J. Phys.: Condens. Matter} \textbf{26}, 473202 (2014).


\bibitem{Boeri2008}
 \textsc{L.~Boeri},  \textsc{O.~Dolgov},  and  \textsc{A.~Golubov},
 \jr{Phys. Rev. Lett.} \textbf{101}, 026403 (2008).


\bibitem{Rettig2013}
 \textsc{L.~Rettig},  \textsc{R.~Cortes},  \textsc{H.~Jeevan},
  \textsc{P.~Gegenwart},  \textsc{T.~Wolf},  \textsc{J.~Fink},  and
  \textsc{U.~Bovensiepen},
 \jr{New J.\ Phys.} \textbf{15}, 083023 (2013).


\bibitem{Hardy2013}
 \textsc{F.~Hardy},  \textsc{R.~Eder},  \textsc{M.~Jackson},  \textsc{A.~Dai},
  \textsc{C.~Paulsen},  \textsc{T.~Wolf},  \textsc{P.~Burger},
  \textsc{A.~Boehmer},  \textsc{P.~Schweiss},  \textsc{P.~Adelmann},
  \textsc{R.~Fisher},  and  \textsc{C.~Meingast},
 \jr{J.\ Phys. Soc.\ Jpn.} \textbf{83}, 014711 (2013).


\bibitem{Hardy2016}
 \textsc{F.~Hardy},  \textsc{A.~B\"ohmer},  \textsc{L.~de~Medici},
  \textsc{M.~Capone},  \textsc{G.~Giovannetti},  \textsc{R.~Eder},
  \textsc{L.~Wang},  \textsc{M.~He},  \textsc{T.~Wolf},  \textsc{P.~Schweiss},
  \textsc{R.~Heid},  \textsc{A.~Herbig},  \textsc{P.~Adelmann},
  \textsc{R.~Fisher},  and  \textsc{C.~Meingast},
 \jr{Phys.\ Rev.\ B} \textbf{94}, 205113 (2016).


\bibitem{Reid2012}
 \textsc{J.\,P. Reid},  \textsc{M.~Tanatar},  \textsc{A.~Juneau-Fecteau},
  \textsc{R.~Gorton},  \textsc{S.~de~Cortret},  \textsc{N.~Doiron-Leyraud},
  \textsc{T.~Saito},  \textsc{H.~Fukuzawa},  \textsc{Y.~Kohoni},
  \textsc{K.~Kihour},  \textsc{C.~Lee},  \textsc{A.~Iyo},  \textsc{H.~Eisaki},
  \textsc{R.~Prozorov},  and  \textsc{L.~Taillefer},
 \jr{Phys.\ Rev.\ Lett.} \textbf{109}, 087001 (2012).


\bibitem{Grinenko2013}
 \textsc{V.~Grinenko},  \textsc{S.\,L. Drechsler},  \textsc{M.~Abdel-Hafiez},
  \textsc{S.~Aswartham},  \textsc{A.~Wolter},  \textsc{S.~Wurmehl},
  \textsc{C.~Hess},  \textsc{K.~Nenkov},  \textsc{G.~Fuchs},
  \textsc{D.~Efremov},  \textsc{B.~Holzapfel},  \textsc{J.~van\,den Brink},
  and  \textsc{B.~Buechner},
 \jr{Phys.\ Stat.\ Sol.\ B} \textbf{250}, 593 (2013).


\bibitem{Kim2014}
 \textsc{H.~Kim},  \textsc{M.~Tanatar},  \textsc{Y.~Liu},  \textsc{Z.~Sims},
  \textsc{C.~Zhang},  \textsc{P.~Dai},  \textsc{T.~Lagasso},  and
  \textsc{R.~Prozorov},
 \jr{Phys. Rev. B} \textbf{89}, 174519 (2014).


\bibitem{Wang2016b}
 \textsc{Q.~Wang},  \textsc{J.~Park},  \textsc{F.~Y.},  \textsc{Y.~Shen},
  \textsc{Y.~Hao},  \textsc{B.~Pam},  \textsc{J.~Lynn},  \textsc{A.~Ivanov},
  \textsc{S.~Chi},  \textsc{M.~Matsuda},  \textsc{H.~Cao},
  \textsc{R.~Birgenau},  \textsc{D.~Efremov},  and  \textsc{J.~Zhao},
 \jr{Phys.\ Rev.\ Lett.} \textbf{116}, 197004 (2016).


\bibitem{Efremov2016}
 \textsc{D.~Efremov},  \textsc{V.~Grinenko},  \textsc{O.~Dolgov},  and
  \textsc{S.\,L. Drechsler},
 \jr{to be published} (2017).


\bibitem{Yin2011}
 \textsc{Z.\,P. Yin},  \textsc{K.~Haule},  and  \textsc{G.~Kotliar},
 \jr{Nature Materials} \textbf{10}, 932 (2011).


\bibitem{deMedici2014}
 \textsc{L.~de~Medici},  \textsc{G.~Giovannetti},  and  \textsc{M.~Capone},
 \jr{Phys.\ Rev.\ Lett.} \textbf{112}, 177001 (2014).


\bibitem{Johnston2014}
 \textsc{S.~Johnston},  \textsc{M.~Abdel-Hafiez},  \textsc{L.~Harnagea},
  \textsc{V.~Grinenko},  \textsc{D.~Bombor},  \textsc{Y.~Krupskaya},
  \textsc{C.~Hess},  \textsc{S.~Wurmehl},  \textsc{A.\,U.\,B. Wolter},
  \textsc{B.~Buechner},  \textsc{H.~Rosner},  and  \textsc{S.\,L.
  Drechsler},
 \jr{Physical Review B} \textbf{89}, 134507 (2014).


\bibitem{Rademaker2016}
 \textsc{L.~Rademaker},  \textsc{Y.~Wang},  \textsc{T.~Berlijn},  and
  \textsc{S.~Johnston},
 \jr{New Journal of Physics} \textbf{18}, 022001 (2016).


\bibitem{WangSUST2016}
 \textsc{Y.~Wang},  \textsc{K.~Nakatsukasa},  \textsc{L.~Rademaker},
  \textsc{T.~Berlijn},  and  \textsc{S.~Johnston},
 \jr{Supercond.\ Sci.\ Technol.} \textbf{29}, 054009 (2016).


\bibitem{Benfatto2009}
 \textsc{L.~Benfatto},  \textsc{E.~Cappelluti},  and
  \textsc{C.~Castellani},
 \jr{Phys.\ Rev.\ B} \textbf{80}, 214522 (2009).


\bibitem{Dolgov2013}
 \textsc{O.~Dolgov},  \textsc{D.~Efremov},  \textsc{M.~Korshunov},
  \textsc{A.~Charnukha},  \textsc{A.~Boris},  and  \textsc{A.~Golubov},
 \jr{J.\ of Supercond.\ and Novel Magnet.} \textbf{26}, 2637--2640 (2013).


\bibitem{Ummarino2015}
 \textsc{G.~Ummarino} and  \textsc{D.~Daghero},
 \jr{J.\ Phys.\ Condens. Mat.} \textbf{27}, 435701 (2015).


\bibitem{Ummarino2009}
 \textsc{G.~Ummarino},  \textsc{M.~Tortello},  \textsc{D.~Daghero},  and
  \textsc{R.~Gonelli},
 \jr{Phys.\ Rev.\ B} \textbf{80}, 172503 (2009).


\bibitem{Ummarino2013}
 \textsc{G.~Ummarino},  \textsc{S.~Galasso},  and  \textsc{A.~Sanna},
 \jr{J.\ Phys.\ Cond.\ Mat.} \textbf{25}, 205701 (2013).


\bibitem{Ummarino2011}
 \textsc{G.~Ummarino},
 \jr{Phys.\ Rev.\ B} \textbf{83}, 092508 (2011).


\bibitem{Drechsler2016}
 \textsc{S.\,L. Drechsler},  \textsc{S.~Borisenko},  \textsc{S.~Khim},
  \textsc{V.~Grinenko},  \textsc{G.~Fuchs},  \textsc{S.~Aswartham},
  \textsc{S.~Wurmehl},  \textsc{B.~B\"uchner},  \textsc{S.~Backes},
  \textsc{R.~Valenti},  \textsc{J.~Tomczak},  \textsc{I.~Morozov},  and
  \textsc{H.~Rosner},
 \jr{Abstracts of the International Workshop on Iron-based superconductors, DFG
  SPP 1458} pp.\,P. 16, Munich, Germany, September 13--16 (2016).


\bibitem{DMFTReview}
 \textsc{A.~van Roekeghem},  \textsc{P.~Richard},  \textsc{H.~Ding},  and
  \textsc{S.~Biermann},
 \jr{Comptes Rendus Physique} \textbf{17}, 140 (2015).


\bibitem{Mazin2008}
 \textsc{I.\,I. Mazin},  \textsc{D.\,J. Singh},  \textsc{M.\,D. Johannes},  and
   \textsc{M.\,H. Du},
 \jr{Phys. Rev. Lett.} \textbf{101}, 057003 (2008).


\bibitem{ScalapinoRMP}
 \textsc{D.\,J. Scalapino},
 \jr{Rev. Mod. Phys.} \textbf{84}, 1383 (2012).


\bibitem{Kontani2010}
 \textsc{H.~Kontani} and  \textsc{S.~Onari},
 \jr{Phys.\ Rev.\ Lett.} \textbf{104}, 157001 (2010).


\bibitem{Kang2016}
 \textsc{J.~Kang} and  \textsc{R.\,M. Fernandes},
 \jr{Phys. Rev. Lett.} \textbf{117}, 217003 (2016).


\bibitem{Capone2010}
 \textsc{M.~Capone},  \textsc{C.~Castellani},  and  \textsc{M.~Grilli},
 \jr{Advances in Condensed Matter Physics} \textbf{2010}, 920860 (2010).


\bibitem{Mandal2014}
 \textsc{S.~Mandal},  \textsc{R.\,E. Cohen},  and  \textsc{K.~Haule},
 \jr{Phys. Rev. B} \textbf{82}, 020506 (2014).


\bibitem{Coh2015}
 \textsc{S.~Coh},  \textsc{M.\,L. Cohen},  and  \textsc{S.\,G. Louie},
 \jr{New J. Phys.} \textbf{17}, 073027 (2015).


\bibitem{Coh2016}
 \textsc{S.~Coh},  \textsc{M.\,L. Cohen},  and  \textsc{S.\,G. Louie},
 \jr{Phys. Rev. B} \textbf{94}, 104505 (2016).


\bibitem{Sawatzky2009}
 \textsc{G.~Sawatzky},  \textsc{I.~Elfimov},  \textsc{J.~van\,den Brink},  and
  \textsc{J.~Zaanen},
 \jr{EPL} \textbf{86}, 17006 (2009).


\bibitem{Drechsler2009}
 \textsc{S.\,L. Drechsler},  \textsc{H.~Rosner},  \textsc{M.~Grobosch},
  \textsc{G.~Behr},  \textsc{F.~Roth},  \textsc{G.~Fuchs},
  \textsc{K.~Koepernik},  \textsc{R.~Schuster},  \textsc{J.~Malek},
  \textsc{S.~Elgazzar},  \textsc{M.~Rotter},  \textsc{D.~Johrendt},
  \textsc{H.\,H. Klauss},  \textsc{B.~Buechner},  and
  \textsc{M.~Knupfer}, arXiv:0904.0827v1 (2009).


\bibitem{Kulic2009a}
 \textsc{M.\,L. Kuli\ifmmode\,\acute{c}\else \'{c}\fi{}} and  \textsc{A.\,A.
  Haghighirad},
 \jr{EPL} \textbf{870} (2009).


\bibitem{Johnston2016a}
 \textsc{S.~Johnston},  \textsc{C.~Monney},  \textsc{V.~Bisogni},
  \textsc{K.\,J. Zhou},  \textsc{R.~Kraus},  \textsc{G.~Behr},
  \textsc{V.~Strocov},  \textsc{J.~Málek},  \textsc{S.\,L. Drechsler},
  \textsc{T.~Geck~J.Schmitt.},  and  \textsc{J.~van\,den Brink},
 \jr{nature comm.} \textbf{7}, 10563 (2016).


\bibitem{Rettig2015}
 \textsc{L.~Rettig},  \textsc{S.~Mariager},  \textsc{A.~Ferrer},
  \textsc{S.~Gr\"ubel},  \textsc{J.~Johnson},  \textsc{J.~Rittmann},
  \textsc{T.~Wolf},  \textsc{S.~Johnson},  \textsc{G.~Ingold},
  \textsc{P.~Beaud},  and  \textsc{U.~Staub},
 \jr{Phys.\ Rev.\ Lett.} \textbf{15}, 067402 (2015).


\bibitem{Cantoni2015}
 \textsc{C.~Cantoni},  \textsc{A.~McGuire},  \textsc{B.~Saparov},
  \textsc{A.~May},  \textsc{T.~Keiber},  \textsc{F.~Bridges},
  \textsc{A.~Sefat},  and  \textsc{B.~Sales},
 \jr{Advanced Materials} \textbf{27}, 2715 (2015).


\bibitem{Barzykin2009}
 \textsc{V.~Barzykin} and  \textsc{L.~Gor'kov},
 \jr{Phys.\ Rev.\ B} \textbf{79}, 134510 (2009).


\bibitem{Widom2016}
 \textsc{M.~Widom} and  \textsc{K.~Quander},
 \jr{J.\ Supercond.\ \& Novel Magnet.} \textbf{29}, 685 (2016).


\bibitem{Kumar2010a}
 \textsc{P.~Kumar},  \textsc{A.~Kumar},  \textsc{S.~Saha},  \textsc{D.~Muthu},
  \textsc{J.~Prakash},  \textsc{U.~Waghmare},  \textsc{A.~Ganguli},  and
  \textsc{A.~Sood},
 \jr{J.\ Phys.\ Condens.\ Mat.} \textbf{22}, 254402 (2010).


\bibitem{Kumar2010b}
 \textsc{A.~Kumar},  \textsc{P.~Kumar},  \textsc{V.~Umesh},
  \textsc{U.~Waghmare},  and  \textsc{A.~Sood},
 \jr{J.\ Phys.\ Condens.\ Mat.} \textbf{22}, 385701 (2010).


\bibitem{Kumar2014}
 \textsc{P.~Kumar},  \textsc{D.~Muthu},  \textsc{L.~Hanargea},
  \textsc{S.~Wurmehl},  \textsc{B.~B\"uchner},  and  \textsc{A.~Sood},
 \jr{J.\ Phys.\ Condens.\ Mat.} \textbf{26}, 305403 (2014).


\bibitem{Zeyher1996}
 \textsc{R.~Zeyher} and  \textsc{M.\,L. Kuli{\'c}},
 \jr{Phys. Rev. B} \textbf{53}, 2850 (1996).


\bibitem{Lee2014}
 \textsc{J.~Lee},  \textsc{F.~Schmitt},  \textsc{R.~Moore},
  \textsc{S.~Johnston},  \textsc{Y.\,T. Cui},  \textsc{W.~Li},  \textsc{M.~Yi},
   \textsc{Z.~Liu},  \textsc{M.~Hashimoto},  \textsc{Y.~Zhang},
  \textsc{D.~Lu},  \textsc{T.~Devereaux},  \textsc{D.\,H. Lee},  and
  \textsc{Z.\,H. Shen},
 \jr{Nature} \textbf{515}, 245 (2014).


\bibitem{Maier2008}
 \textsc{T.\,A. Maier},  \textsc{D.~Poiblanc},  and  \textsc{D.\,J. Scalapino},
 \jr{Phys. Rev. Lett.} \textbf{100}, 237001 (2008).


\bibitem{Iwasawa2010}
 \textsc{H.~Iwasawa},  \textsc{Y.~Yoshida},  \textsc{I.~Hase},
  \textsc{S.~Koikegami},  \textsc{H.~Hyashi},  \textsc{J.~Jiang},
  \textsc{K.~Shimada},  \textsc{M.~Hirano},  \textsc{H.~Namatame},
  \textsc{M.~Taniguchi},  and  \textsc{Y.~Aiura},
 \jr{Phys.\ Rev.\ Lett.} \textbf{105}, 226406 (2010).


\bibitem{Iwasawa2013}
 \textsc{H.~Iwasawa},  \textsc{Y.~Yoshida},  \textsc{M.~Hirano},
  \textsc{I.~Hase},  \textsc{K.~Shimada},  \textsc{H.~Namatame},
  \textsc{M.~Taniguchi},  and  \textsc{Y.~Aiura},
 \jr{Scient.\ Rep.} \textbf{3}, 1930 (2013).


\bibitem{Bulut1996}
 \textsc{N.~Bulut} and  \textsc{D.\,J. Scalapino},
 \jr{Phys.\ Rev.\ B} \textbf{54}, 14971 (1996).


\bibitem{JohnstonDOS}
 \textsc{S.~Johnston} and  \textsc{T.\,P. Devereaux},
 \jr{Phys. Rev. B} \textbf{81}, 214512 (2010).


\bibitem{Dai2016}
 \textsc{Y.~Dai},  \textsc{H.~Miao},  \textsc{L.~Xing},  \textsc{X.~Wang},
  \textsc{C.~Jin},  \textsc{H.~Ding},  and  \textsc{C.~Homes},
 \jr{Phys.\ Rev.\ B} \textbf{93}, 054508 (2016).


\bibitem{Cheng2012}
 \textsc{B.~Cheng},  \textsc{B.\,F. Hu},  \textsc{R.\,Y. Chen},
  \textsc{G.~Xu},  \textsc{P.~Zheng},  \textsc{J.\,L. Luo},  and
  \textsc{N.\,L. Wang},
 \jr{Phys.\ Rev.\ B} \textbf{86}, 134503 (2012).


\bibitem{Drechsler2010}
 \textsc{S.\,L. Drechsler},  \textsc{F.~Roth},  \textsc{M.~Grobosch},
  \textsc{R.~Schuster},  \textsc{K.~Koepernik},  \textsc{H.~Rosner},
  \textsc{G.~Behr},  \textsc{M.~Rotter},  \textsc{D.~Johrendt},
  \textsc{B.~Buechner},  and  \textsc{M.~Knupfer},
 \jr{Physica C} \textbf{470}, S332 (2010).


\bibitem{Maksimov2011}
 \textsc{E.~Maksimov},  \textsc{A.~Karakozov},  \textsc{B.~Gorshunova},
  \textsc{Y.~Ponomaeb},  and  \textsc{M.~Dressel},
 \jr{Phys.\ Rev.\ B} \textbf{84}, 174504 (2011).


\bibitem{Wu2010}
 \textsc{Y.~Wu},  \textsc{N.~Barisi{\'c}},  \textsc{P.~Kallina},
  \textsc{A.~Faridian},  \textsc{B.~Gorshunv},  \textsc{N.~Drichko},
  \textsc{L.\,J. Li},  \textsc{X.~Lin},  \textsc{G.\,H. Cao},  \textsc{Z.\,A.
  Xu},  \textsc{N.\,L. Wang},  and  \textsc{M.~Dressel},
 \jr{Phys.\ Rev. B} \textbf{81}, 100512(R) (2010).


\bibitem{Charnukha2013}
 \textsc{A.~Charnukha},  \textsc{D.~Pr\"opper},  \textsc{T.~Larkin},
  \textsc{D.~Sun},  \textsc{Z.~Li},  \textsc{C.~Lin},  \textsc{T.~Wolf},
  \textsc{B.~Keimer},  and  \textsc{A.~Boris},
 \jr{Phys.\ Rev.\ B} \textbf{88}, 184511 (2013).


\bibitem{Dai2016a}
 \textsc{Y.~Dai},  \textsc{A.~Akrap},  \textsc{L.~Bud'ko},
  \textsc{P.~Canfield},  and  \textsc{C.~Homes},
 \jr{Phys.\ Rev.\ B} \textbf{94}, 195142 (2016).


\bibitem{Tytarenko2015}
 \textsc{A.~Tytarenko},  \textsc{Y.~Huang},  \textsc{A.~de~Visser},
  \textsc{S.~Johnston},  and  \textsc{E.~van Heumen},
 \jr{Sci.\ Rep.} \textbf{5}, 12421 (2015).


\bibitem{vanHeumenEPL}
 \textsc{E.~van Heumen},  \textsc{Y.~Huang},  \textsc{S.~de~Jong},
  \textsc{A.\,B. Kuzmenko},  \textsc{M.\,S. Golden},  and  \textsc{D.~van\,der
  Marel},
 \jr{Europhys.\ Lett.} \textbf{90}, 37005 (2010).


\bibitem{BenfattoInterband}
 \textsc{L.~Benfatto},  \textsc{E.~Cappelluti},  \textsc{L.~Ortenzi},  and
  \textsc{L.~Boeri},
 \jr{Phys. Rev. B} \textbf{83}, 224514 (2011).


\bibitem{Diel2014}
 \textsc{J.~Diel},  \textsc{S.~Backes},  \textsc{D.~Guterding},
  \textsc{H.~Jeschke},  and  \textsc{V.~Roser},
 \jr{Phys.\ Rev.\ B} \textbf{90}, 085110 (2014).


\bibitem{Materne2015}
 \textsc{P.~Materne},  \textsc{S.~Kamusella},  \textsc{R.~Sarkar},
  \textsc{T.~Goltz},  \textsc{J.~Spehling},  \textsc{M.~Maeter},
  \textsc{L.~Harnagea},  \textsc{S.~Wurmehl},  \textsc{B.~Buechner},
  \textsc{C.~LuetkensH.~Timm.},  and  \textsc{H.\,H. Klauss},
 \jr{Phys.\ Rev.\ B} \textbf{92}, 134511 (2015).


\bibitem{Drechsler2008}
 \textsc{S.\,L. Drechsler},  \textsc{M.~Grobosch},  \textsc{K.~Koepernik},
  \textsc{G.~Behr},  \textsc{A.~Koehler},  \textsc{J.~Werner},
  \textsc{A.~Kondrat},  \textsc{N.~Leps},  \textsc{C.~Hess},
  \textsc{R.~Klingeler},  \textsc{R.~Schuster},  \textsc{B.~Buechner},  and
  \textsc{M.~Knupfer},
 \jr{Phys.\ Rev.\ Lett.} \textbf{101}, 257004 (2008).


\bibitem{Guguchia2016}
 \textsc{Z.~Guguchia},  \textsc{R.~Khasanov},  \textsc{Z.~Bukowski},
  \textsc{F.~von Rohr},  \textsc{M.~Medarde},  \textsc{P.~Biswas},
  \textsc{H.~Luetkens},  \textsc{A.~Amato},  and  \textsc{E.~Morenzoni},
 \jr{Phys.\ Rev.\ B} \textbf{93}, 094513 (2016).


\bibitem{HundsMetal}
 \textsc{A.~Georges},  \textsc{L.~de' Medici},  and  \textsc{J.~Mravlje},
 \jr{Annual Reviews of Condensed Matter Physics} \textbf{4}, 137 (2013).


\bibitem{Medici2016}
 \textsc{L.~de~Medici} and  \textsc{M.~Capone}, arXiv:1607.08468 (2016).


\bibitem{Brouet2016}
 \textsc{V.~Brouet},  \textsc{P.~LeBeuf},  \textsc{P.\,H. Lin},
  \textsc{J.~Mansart},  \textsc{A.~Taleb-Ibrahimi},  \textsc{P.~Le~Fevre},
  \textsc{F.~Bertrand},  \textsc{A.~Forget},  and  \textsc{D.~Colson},
 \jr{Phys.\ Rev.\ B} \textbf{93}, 085137 (2016).


\bibitem{Roekeghem2016}
 \textsc{A.~Roekeghem},  \textsc{L.~Vaugier},  \textsc{H.~Jiang},  and
  \textsc{S.~Biermann},
 \jr{Phys.\ Rev.\ B} \textbf{94}, 125147 (2016).


\bibitem{Guterding2016}
 \textsc{D.~Guterding},  \textsc{S.~Backes},  \textsc{M.~Tomic},
  \textsc{H.~Jeschke},  and  \textsc{R.~Valenti},
 \jr{phys.\ stat.\ sol.\ (b) (in press)}, arXiv:1606.04411v1 (2016).


\bibitem{Backes2015}
 \textsc{S.~Backes},  \textsc{M.~Tomic},  \textsc{H.~Jeschke},  and
  \textsc{R.~Valenti},
 \jr{Phys.\ Rev.\ B} \textbf{92}, 195128 (2015).


\bibitem{Carbotte1990}
 \textsc{J.~Carbotte},
 \jr{Rev.\ Mod.\ Phys.} \textbf{62}, 1027 (1990).


\bibitem{Anderson2007}
 \textsc{P.\,W. Anderson},
 \jr{Science} \textbf{316}, 1705 (2007).


\bibitem{Fuchs2008}
 \textsc{G.~Fuchs},  \textsc{S.\,.\,L. Drechsler},  \textsc{N.~Kozlova},
  \textsc{G.~Behr},  \textsc{A.~Koehler},  \textsc{J.~Werner},
  \textsc{K.~Nenkov},  \textsc{R.~Klingeler},  \textsc{J.~Hamann-Borrero},
  \textsc{C.~Hess},  \textsc{A.~Kondrat},  \textsc{M.~Grobosch},
  \textsc{A.~Narduzzo},  \textsc{M.~Knupfer},  \textsc{J.~Freudenberger},
  \textsc{B.~Buechner},  and  \textsc{L.~Schultz},
 \jr{Phys.\ Rev.\ Lett.} \textbf{101}, 237003 (2008).


\bibitem{Fuchs2009}
 \textsc{G.~Fuchs},  \textsc{S.\,L. Drechsler},  \textsc{N.~Kozlova},
  \textsc{M.~Bartkowiak},  \textsc{J.\,E. Hamann-Borrero},  \textsc{G.~Behr},
  \textsc{K.~Nenkov},  \textsc{H.\,H. Klauss},  \textsc{H.~Maeter},
  \textsc{A.~Amato},  \textsc{H.~Luetkens},  \textsc{A.~Kwadrin},
  \textsc{R.~Khasanov},  \textsc{J.~Freudenberger},  \textsc{A.~Koehler},
  \textsc{M.~Knupfer},  \textsc{E.~Arushanov},  \textsc{H.~Rosner},
  \textsc{B.~Buechner},  and  \textsc{L.~Schultz},
 \jr{New Journal of Physics} \textbf{11}, 075007 (2009).


\bibitem{Charnukha2011}
 \textsc{A.~Charnukha},  \textsc{O.~Dolgov},  \textsc{A.~Golubov},
  \textsc{Y.~Matiks},  \textsc{D.~Sun},  \textsc{C.~Lin},  \textsc{B.~Keimer},
  and  \textsc{A.~Boris},
 \jr{Phys.\ Rev.\ B} \textbf{84}, 174511 (2011).


\bibitem{Popovich2010}
 \textsc{P.~Popovich},  \textsc{A.\,V. Boris},  \textsc{O.\,V. Dolgov},
  \textsc{A.\,A. Golubov},  \textsc{D.\,L. Sun},  \textsc{C.\,T. Lin},
  \textsc{R.\,K. Kremer},  and  \textsc{B.~Keimer},
 \jr{Phys.\ Rev.\ Lett.} \textbf{105}, 027003 (2010).


\bibitem{Evtushinsky2013}
 \textsc{D.~Evtushinsky},  \textsc{V.~Zabolotnyy},  \textsc{L.~Harnagea},
  \textsc{A.~Yaresko},  \textsc{S.~Thirupathaiah},  \textsc{A.~Kordyuk},
  \textsc{J.~Maletz},  \textsc{S.~Aswartham},  \textsc{S.~Wurmehl},
  \textsc{E.~Rienks},  \textsc{R.~Follath},  \textsc{B.~Buechner},  and
  \textsc{B.~S.V.},
 \jr{Phys.\ Rev.\ B} \textbf{87}, 094501 (2013).


\bibitem{Shi2014}
 \textsc{Y.\,B. Shi},  \textsc{Y.\,B. Huang},  \textsc{X.\,P. Wang},
  \textsc{X.~Shi},  \textsc{A.\,V. Roekeghem},  \textsc{W.\,L. Zhang},
  \textsc{N.~Xu},  \textsc{P.~Richard},  \textsc{Q.~T.},  \textsc{E.~Rienks},
  \textsc{S.~Thirupathaiah},  \textsc{K.~Zhao},  \textsc{C.\,Q. Jing},
  \textsc{M.~Shi},  and  \textsc{H.~Ding},
 \jr{Chin.\ Phys.\ Lett.} \textbf{31}, 067403 (2014).


\bibitem{Hwang2016}
 \textsc{J.~Hwang},
 \jr{J.\ Phys.\ Condens. Matter} \textbf{28}, 125709 (2016).


\bibitem{Derondeau2016}
 \textsc{G.~Derondeau},  \textsc{F.~Bisti},  \textsc{J.~Braun},
  \textsc{V.~Rogalev},  \textsc{M.~Shi},  \textsc{T.~Schmitt},  \textsc{J.~Ma},
   \textsc{H.~Ding},  \textsc{H.~Ebert},  \textsc{V.~Strocov},  and
  \textsc{J.~Minar},
 \jr{arXiv:1606.08977v1} (2016).


\bibitem{Johnston2016}
 \textsc{S.~Johnston} and  \textsc{{\it et al.}},
 \jr{in preparation} (2016).


\bibitem{Hlobil2016}
 \textsc{P.~Hlobil},  \textsc{J.~Jandke},  \textsc{W.~Wulfhekel},  and
  \textsc{J.~Schmalian}, arXiv:1603.05288v2 (2016).


\othercit
\bibitem{Schmalianprivate2016}
 \textsc{J.~Schmalian},
private comm.\ (2016).


\bibitem{Inosov2011}
 \textsc{D.~Inosov},  \textsc{J.~Park},  \textsc{A.~Charnukha},
  \textsc{Y.~Li},  \textsc{A.~Boris},  \textsc{B.~Keimer},  and
  \textsc{V.~V.~Hinkov},
 \jr{Phys.\ Rev.\ B} \textbf{83}, 14520 (2011).


\bibitem{Kant2010}
 \textsc{C.~Kant},  \textsc{J.~Deisenhofer},  \textsc{A.~G{\"u}nther},
  \textsc{F.~Schrettle},  \textsc{M.~Rotter},  \textsc{D.~Johrendt},  and
  \textsc{A.~Loidl},
 \jr{Phys. Rev. B} \textbf{81}, 014529 (2009).


\bibitem{Ma2010}
 \textsc{F.~Ma},  \textsc{Z.\,Y. Lu},  and  \textsc{T.~Xiang},
 \jr{Front. Phys. China} \textbf{5}, 150 (2010).


\bibitem{Tortello2010}
 \textsc{M.~Tortello},  \textsc{D.~Daghero},  \textsc{G.~Ummarino},
  \textsc{V.~Stepanov},  \textsc{J.~Jiang},  \textsc{J.~Wiss},
  \textsc{E.~Hellstrom},  and  \textsc{R.~Gonelli},
 \jr{Phys.\ Rev.\ Lett.} \textbf{105}, 237002 (2010).


\bibitem{Ortenzi2009}
 \textsc{L.~Ortenzi},  \textsc{E.~Cappelluti},  \textsc{L.~Benfatto},  and
  \textsc{L.~Pietronero},
 \jr{Phys. Rev. Lett.} \textbf{103}, 046404 (2009).


\bibitem{Benfatto2011}
 \textsc{L.~Benfatto} and  \textsc{E.~Cappelluti},
 \jr{Phys. Rev. B} \textbf{83}, 104516 (2011).


\bibitem{Linscheid2016}
 \textsc{A.~Linscheid},  \textsc{S.~Maiti},  \textsc{Y.~Wang},
  \textsc{S.~Johnston},  and  \textsc{P.~Hirschfeld},
 \jr{arXiv:1603.03739v1} (2016).


\bibitem{Chubukov2016}
 \textsc{A.~Chubukov},  \textsc{I.~Eremin},  and  \textsc{D.~Efremov},
 \jr{arXiv:1601.01678v1} (2016).


\bibitem{Valentinis2016}
 \textsc{D.~Valentinis},  \textsc{D.~van\,der Marel},  and
  \textsc{C.~C.~Berthod},
 \jr{arXiv:1601.04521v2} (2016).


\bibitem{Karakozov2014}
 \textsc{A.~Karakozov},  \textsc{S.~Zapf},  \textsc{B.~Gorshunova},
  \textsc{Y.~Ponomarev},  \textsc{M.~Magnitskaya},  \textsc{E.~Zhukova},
  \textsc{A.~Prokhorov},  \textsc{V.~Anzin},  and  \textsc{S.~Hayndl},
 \jr{Phys.\ Rev.\ B} \textbf{90}, 014506 (2014).


\bibitem{Kawaguchi2016}
 \textsc{N.~Kawaguchi},  \textsc{S.~Fujiwara},  \textsc{S.~Iimura},
  \textsc{S.~Matsuishi},  and  \textsc{H.~Hosonol}, arXiv:1609.04957v1
  (2016).


\othercit
\bibitem{Maksimov1982}
 \textsc{V.~Ginzburg} and  \textsc{D.~Kirzhnitz} (eds.),
Maksimov, E.G. and Karakozov, A.E. in {\it "High Temperature
  Superconductivity"} Chapt.\ 7 (New York: Consultant Bureau, 1982).


\othercit
\bibitem{Fulde1978}
 \textsc{P.~Fulde},
in {\it "Handbook on the Physics and Chemistry of Rare Earths"} ed.\ by
  Gscheidner Jr., K.A., vol.\ 2, p. 295 (North-Holland, Amsterdam, 1978).


\bibitem{Naidyuk2007}
 \textsc{Y.~Naidyuk},  \textsc{O.~Kvitnitskaya},  \textsc{D.~Yanson},
  \textsc{G.~Fuchs},  \textsc{K.~Nenkov},  \textsc{A.~W\"alte},
  \textsc{G.~Behr},  \textsc{D.~Souptel},  and  \textsc{S.\,L.
  Drechsler},
 \jr{Phys. Rev. B} \textbf{76}, 014520 (2007).


\bibitem{Xiao2013}
 \textsc{Y.~Xiao},  \textsc{M.~Zbiri},  \textsc{R.~Downie},  \textsc{J.\,W.\,G.
  Bos},  \textsc{T.~Br\"uckel},  and  \textsc{T.~Chatterji},
 \jr{Phys.\ Rev.\ B} \textbf{88}, 214419 (2013).


\bibitem{WangFeSeSTO}
 \textsc{Q.\,Y.\,W. {\it et al.}},
 \jr{Chin. Phys. Lett.} \textbf{29}, 037402 (2012).


\bibitem{Wang2016}
 \textsc{H.~Wang},  \textsc{A.~Kreisel},  \textsc{P.~Hirschfeld},  and
  \textsc{V.~Mishra}, arXiv:1606.02198v1 (2016).


\bibitem{Kyung2016}
 \textsc{W.~Kyung},  \textsc{S.~Huh},  \textsc{Y.~Koh},  \textsc{K.\,Y. Choi},
  \textsc{M.~Nakajima},  \textsc{H.~Eisaki},  \textsc{J.~Denlinger},
  \textsc{S.\,K. Mo},  \textsc{C.~Kim},  and  \textsc{Y.~Kim},
 \jr{Nature Materials} \textbf{15}, 1233 (2016).


\bibitem{Choi2016}
 \textsc{S.~Choi},  \textsc{W.\,J. Jangi},  \textsc{H.\,J. Lee},
  \textsc{J.~Ok},  \textsc{H.~Choi},  \textsc{A.~Lee},  \textsc{A.~Akbari},
  \textsc{K.~Nakatsukasa},  \textsc{Y.~Semertzidis},  \textsc{Y.~Bang},
  \textsc{S.~Johnston},  \textsc{J.~Kim},  and  \textsc{J.~Lee},
 \jr{arXiv:1608.00886v2} (2016).


\bibitem{Kulic2016}
 \textsc{M.\,L. Kuli\ifmmode\,\acute{c}\else \'{c}\fi{}} and
  \textsc{O.~Dolgov}, arXiv:1607.00843 (2016).


\bibitem{Inada1992}
 \textsc{Y.~Inada} and  \textsc{K.~Nasu},
 \jr{J. Phys. Soc. Jpn.} \textbf{61}, 4511 (1992).


\bibitem{Freericks1996}
 \textsc{J.\,K. Freericks},  \textsc{M.~Jarrell},  and  \textsc{G.\,D. Mahan},
 \jr{Phys. Rev. Lett} \textbf{77}, 4588 (1996).


\bibitem{Chang2009}
 \textsc{J.~Chang},  \textsc{I.~Eremin},  and  \textsc{P.~Thalmeier},
 \jr{New Jour. of Phys.} \textbf{11}, 055068 (2009).


\bibitem{Bergmann1973}
 \textsc{L.~Bergmann},  \textsc{E.~Cappelluti},  and
  \textsc{C.~Castellani},
 \jr{Phys.\ Rev.\ B} \textbf{7}, 480 (2009).


\bibitem{Rainer1973}
 \textsc{D.~Rainer},  \textsc{G.~Bergmann},  and  \textsc{U.~Eckhardt},
 \jr{Phys.\ Rev.\ B} \textbf{8}, 5324 (1973).


\bibitem{Ivanov2016}
 \textsc{V.~Ivanov},  \textsc{A.~Ivanov},  \textsc{A.\,P. Menushenkov},
  \textsc{B.~Joseph},  and  \textsc{A.~Bianconi},
 \jr{J. Supercond. \ Nov. Mat.} \textbf{29}, 3035 (2016).


\bibitem{Eilers2016}
 \textsc{F.~Eilers},  \textsc{K.~Grube},  \textsc{D.~Zocco},  \textsc{T.~Wolf},
   \textsc{M.~Merz},  \textsc{P.~Schweiss},  \textsc{R.~Heid},
  \textsc{R.~Eder},  \textsc{R.~Yu},  \textsc{J.\,X. Zhu},  \textsc{Q.~Si},
  \textsc{T.~Shibauchi},  and  \textsc{H.~Hilbert\,v. L\"ohneysen},
 \jr{Phys. Rev. Lett.} \textbf{116}, 237003 (2016).


\bibitem{Lee2011}
 \textsc{C.\,H. Lee},  \textsc{K.~Kihou},  \textsc{H.~Kawano-Furukawa},
  \textsc{T.~Saito},  \textsc{A.~Iyo},  \textsc{H.~Eisaki},
  \textsc{H.~Fukazawa},  \textsc{Y.~Kohori},  \textsc{K.~Suzuki},
  \textsc{H.~Usui},  \textsc{K.~Kuroki},  and  \textsc{K.~Yamada},
 \jr{Phys.\ Rev.\ Lett.} \textbf{116}, 067003 (2011).


\bibitem{Zingl2016}
 \textsc{M.~Zingl},  \textsc{E.~Assmann},  \textsc{P.~Seth},
  \textsc{I.~Krivenko},  and  \textsc{M.~Aichhorn},
 \jr{Phys.\ Rev.\ B} \textbf{94}, 045130 (2016).


\bibitem{Wu2016}
 \textsc{Y.~Wu},  \textsc{D.~Zhao},  \textsc{A.~Wang},  \textsc{N.~Wang},
  \textsc{Z.~Xiang},  \textsc{X.~Luo},  \textsc{T.~Wu},  and
  \textsc{X.~Chen},
 \jr{Phys.\ Rev.\ Lett.} \textbf{116}, 147001 (2016).


\bibitem{fplo}
 \textsc{K.~Koepernik} and  \textsc{H.~Eschrig},
 \jr{Phys. Rev. B} \textbf{59}, 1743 (1999).


\bibitem{fplo2}
 \textsc{I.~Opahle},  \textsc{K.~Koepernik},  and  \textsc{H.~Eschrig},
 \jr{Phys.\ Rev.\ B} \textbf{60}, 14035 (1999).


\bibitem{PBE}
 \textsc{J.\,P. Perdew},  \textsc{B.~K.},  and  \textsc{M.~Ernzerhof},
 \jr{Phys.\ Rev.\ Lett} \textbf{77}, 3865 (1996).


\bibitem{KSD}
 \textsc{H.~Eschrig},  \textsc{M.~Richter},  and  \textsc{I.~Opahle},
 \jr{Theoretical and Computational Chemistry} \textbf{13}, 733 (2004).


\bibitem{Zhang2015}
 \textsc{Z.~Wang},  \textsc{A.~Wang},  \textsc{X.~Hong},  \textsc{J.~Zhang},
  \textsc{B.~Pan},  \textsc{P.~Pan},  \textsc{Y.~Xu},  \textsc{X.~Luo},
  \textsc{X.~Chen},  and  \textsc{S.~Li},
 \jr{Phys.\ Rev.\ B} \textbf{91}, 024502 (2015).


\bibitem{Khim2016}
 \textsc{S.~Khim},  \textsc{S.~Aswartham},  \textsc{V.~Grinenko},
  \textsc{D.~Efremov},  \textsc{C.~Blum},  \textsc{F.~Steckel},
  \textsc{D.~Gruner},  \textsc{A.~Wolter},  \textsc{S.\,L. Drechsler},
  \textsc{C.~Hess},  \textsc{S.~Wurmehl},  and  \textsc{B.~Buechner},
 \jr{phys.\ stat.\ sol.\ (b)} \textbf{xxx}, in press (2016).


\bibitem{Mizukami2016}
 \textsc{Y.~Mizukami},  \textsc{Y.~Kawamoto},  \textsc{Y.~Shimoyama},
  \textsc{S.~Kurata},  \textsc{H.~Ikeda},  \textsc{T.~Wolf},
  \textsc{D.~Zocco},  \textsc{K.~Grube},  and  \textsc{H.~L\"ohneysen},
 \jr{Phys.\ Rev.\ B} \textbf{94}, 024508 (2016).


\bibitem{Terashima2009}
 \textsc{T.~Terashima},  \textsc{M.~Kimata},  \textsc{H.~Satsukawa},
  \textsc{A.~Harada},  \textsc{K.~Hazama},  \textsc{S.~Uji},
  \textsc{H.~Harima},  \textsc{G.\,F. Chen},  \textsc{J.\,L. Luo},  and
  \textsc{N.\,L. Wang},
 \jr{J.\ Phys.\ Soc.\ Jpn.} \textbf{78}, 063702 (2009).


\othercit
\bibitem{Eilers2014}
 \textsc{F.~Eilers},
PhD-Thesis: Superconductivity and electronic structure of KFe2As2, RbFe2As2,
  and CsFe2As2 probed by thermal expansion and magnetostriction at very low
  temperatures (University Press, Stuttgart, 2014).


\bibitem{Vafek2016}
 \textsc{O.~Vafek} and  \textsc{A.~Chubukov},
 \jr{arXiv:1611.05802v1} (2016).


\bibitem{Grinenko2014a}
 \textsc{V.~Grinenko},  \textsc{D.\,V. Efremov},  \textsc{S.\,L. Drechsler},
  \textsc{S.~Aswartham},  \textsc{D.~Gruner},  \textsc{M.~Roslova},
  \textsc{I.~Morozov},  \textsc{K.~Nenkov},  \textsc{S.~Wurmehl},
  \textsc{A.~Wolter},  \textsc{B.~Holzapfel},  and
  \textsc{B.~Buechner},
 \jr{Phys.\ Rev.\ B} \textbf{89}, 060504 (2014).


\bibitem{Abdel-Hafiez2013}
 \textsc{M.~Abdel-Hafiez},  \textsc{V.~Grinenko},  \textsc{S.~Aswartham},
  \textsc{I.~Morozov},  \textsc{M.~Roslova},  \textsc{O.~Vakaliuk},
  \textsc{S.~Johnston},  \textsc{D.\,V. Efremov},  \textsc{J.~van\,den Brink},
  \textsc{H.~Rosner},  \textsc{M.~Kumar},  \textsc{C.~Hess},
  \textsc{S.~Wurmehl},  \textsc{A.~Wolter},  \textsc{B.~Buechner},
  \textsc{E.~Green},  \textsc{J.~Wosnitza},  \textsc{P.~Vogt},
  \textsc{A.~Reifenberger},  \textsc{C.~Enss},  \textsc{M.~Hempel},
  \textsc{R.~Klingeler},  and  \textsc{S.\,L. Drechsler},
 \jr{ibid.} \textbf{87}, 180507 (2013).


\bibitem{Okazaki2012}
 \textsc{K.~Okazaki},  \textsc{Y.~Ota},  \textsc{Y.~Kotani},
  \textsc{W.~Malaeb},  \textsc{Y.~Ishida},  \textsc{S.~T.},  \textsc{T.~Kiss},
  \textsc{W.~S.},  \textsc{C.\,T. Chen},  \textsc{K.\,. Kihou},  and
  \textsc{{\it et}.~{\it al.}}


\bibitem{Ota2014}
 \textsc{Y.~Ota},  \textsc{K.~Okazaki},  \textsc{Y.~Kotani},
  \textsc{T.~Shimojima},  \textsc{W.~Malaeb},  \textsc{S.~Watanabe},
  \textsc{C.\,T. Chen},  \textsc{K.~Kihou},  \textsc{C.~Lee},  \textsc{A.~Iyo},
   and  \textsc{et~al.},
 \jr{Phys.\ Rev.\ B} \textbf{89}, 081103 (2014).


\bibitem{Amano2015}
 \textsc{Y.~Amano},  \textsc{M.~Ishihara},  \textsc{M.~Ichioka},
  \textsc{N.~Nakai},  and  \textsc{K.~Machida},
 \jr{Phys.\ Rev.\ B} \textbf{91}, 144513 (2015).


\bibitem{Grinenko2014}
 \textsc{V.~Grinenko},  \textsc{W.~Schottenhamel},  \textsc{A.\,U.\,B. Wolter},
   \textsc{D.\,V. Efremov},  \textsc{S.\,.\,L. Drechsler},
  \textsc{S.~Aswartham},  \textsc{M.~Kumar},  \textsc{S.~Wurmehl},
  \textsc{M.~Roslova},  \textsc{I.\,V. Morozov},  \textsc{B.~Holzapfel},
  \textsc{B.~Buechner},  \textsc{E.~Ahrens},  \textsc{S.\,I. Troyanov},
  \textsc{S.~Koehler},  \textsc{E.~Gati},  \textsc{S.~Knoener},  \textsc{N.\,H.
  Hoang},  \textsc{M.~Lang},  \textsc{F.~Ricci},  and
  \textsc{G.~Profeta},
 \jr{Physical Review B} \textbf{90}, 094511 (2014).


\bibitem{Yoshida2014}
 \textsc{T.~Yoshida},  \textsc{S.~S.~Ideta},  \textsc{I.~Nishi},
  \textsc{A.~Fujimori},  \textsc{M.~Yi},  \textsc{S.~MooreR.G.~Mo.},
  \textsc{D.\,H. Lu},  \textsc{S.\,Z. Shen},  \textsc{Z.~Hussain},
  \textsc{K.~Kihou},  \textsc{P.~Shirage},  \textsc{H.~Kito},  \textsc{C.\,H.
  Lee},  \textsc{A.~Iyo},  \textsc{H.~Eisaki},  and  \textsc{H.~Harima},
 \jr{Front.\ Phys.} \textbf{2}, 17 (2014).


\bibitem{Naidyuk2015}
 \textsc{Y.~Naidyuk},  \textsc{O.\,E. Kvitnitskaya},  \textsc{N.~Gamayunova},
  \textsc{L.~Boeri},  \textsc{S.~Aswartham},  \textsc{S.~Wurmehl},
  \textsc{B.~B\"uchner},  \textsc{D.~Efremov},  \textsc{G.~Fuchs},  and
  \textsc{S.\,L. Drechsler},
 \jr{Phys. Rev. B} \textbf{90}, 094505 (2014).


\bibitem{Avci2012}
 \textsc{S.~Avci},  \textsc{O.~Chmaissem},  \textsc{D.~Ghung},
  \textsc{S.~Rosenkrans},  \textsc{E.~Goremychkin},  \textsc{J.~Castellan},
  \textsc{I.~Todorov},  \textsc{J.~Schlueter},  \textsc{H.~Claus},
  \textsc{A.~Daoud-Aladine},  \textsc{D.~Khalyavin},  \textsc{M.~Kanatsidis},
  and  \textsc{R.~Osborn},
 \jr{Phys.\ Rev.\ B} \textbf{85}, 184507 (2012).


\bibitem{Wang2016a}
 \textsc{P.\,S. Wang},  \textsc{P.~Zhoul},  \textsc{J.~Dai},
  \textsc{J.~Zhang},  \textsc{X.\,X. Ding},  \textsc{P.~Lin},  \textsc{H.~Wen},
   \textsc{B.~Normand},  \textsc{R.~Yu},  and  \textsc{W.~Yu},
 \jr{Phys.\ Rev.\ B} \textbf{93}, 085129 (2016).


\bibitem{Civardi2016}
 \textsc{E.~Civardi},  \textsc{M.~Moroni},  \textsc{M.~Babij},
  \textsc{X.~Bukowsky},  and  \textsc{P.~Caretta},
 \jr{arXiv:04532v1} (2016).


\bibitem{Berciu}
 \textsc{C.\,P.\,J. Adolphs} and  \textsc{M.~Berciu},
 \jr{Phys. Rev. B} \textbf{90}, 085149 (2014).


\bibitem{Kuroki2010}
 \textsc{K.~Kuroki},
 \jr{Proceedings SNS2010}, arXiv:1008.2286 (2010).


\bibitem{Mizuguchi2010}
 \textsc{Y.~Mizuguchi},  \textsc{Y.~Hara},  \textsc{K.~Deguchi},
  \textsc{S.~S~Tsuda},  \textsc{T.~Yamaguchi},  \textsc{K.~Takeda},
  \textsc{H.~Kotegawa},  \textsc{H.~Tou},  and  \textsc{Y.~Takano},
 \jr{Supercond.\ Sc.\ and Techn.} \textbf{5}, 054013 (2010).


\bibitem{Fang2015}
 \textsc{D.~Fang},  \textsc{X.~Shi},  \textsc{Z.~Du},  \textsc{P.~Richard},
  \textsc{H.~Yang},  \textsc{X.~Wu},  \textsc{P.~Zhang},  \textsc{T.~Qian.},
  \textsc{X.~Ding},  \textsc{Z.~Wang},  \textsc{T.~Kim},  \textsc{M.~Hoesch},
  \textsc{A.~Wang},  \textsc{X.~Chen},  \textsc{J.~Hu},  \textsc{H.~Ding},  and
   \textsc{H.\,H. Wen},
 \jr{Phys. Rev. B} \textbf{92}, 144513 (2015).


\bibitem{Fernandes2016}
 \textsc{R.~Fernandes} and  \textsc{A.~Chubukov}, arXiv:1607.00865v1
  (2016).


\bibitem{Yang2016}
 \textsc{H.~Yang},  \textsc{J.~Xing},  \textsc{Z.~Du},  \textsc{X.~Yang},
  \textsc{H.~Lin},  \textsc{D.~Fang},  \textsc{X.~Zhu},  and  \textsc{H.\,H.
  Wen},
 \jr{Phys.\ Rev.\ B} \textbf{88}, 224516 (2012).


\bibitem{Zaanen2009}
 \textsc{J.~Zaanen},
 \jr{Phys.\ Rev.\ B} \textbf{80}, 212502 (2009).


\bibitem{Alloul2016}
 \textsc{H.~Alloul},  \textsc{P.~Wzietek},  \textsc{T.~Mito},
  \textsc{D.~Pontiroli},  \textsc{M.~Aramini},  \textsc{M.~Rico},  and
  \textsc{E.~Elkaim}, arXiv:1610.00513 (2016).


\bibitem{Dong2010}
 \textsc{J.~Dong},  \textsc{S.~Zhou},  \textsc{T.~Guan},  \textsc{H.~Zhang},
  \textsc{Y.~Dai},  \textsc{X.~Qiu},  \textsc{X.~Wang},  \textsc{Y.~He},
  \textsc{X.~Chen},  and  \textsc{S.~Li},
 \jr{ibid.} \textbf{104}, 087005 (2010).


\othercit
\bibitem{Radousky2000}
 \textsc{H.~Radousky},
Magnetism in heavy fermion compounds, 2000.


\bibitem{Drechsler1997}
 \textsc{S.\,L. Drechsler},  \textsc{J.~Malek},  \textsc{H.~Eschrig},
  \textsc{H.~Rosnerr},  and  \textsc{R.~Hayn},
 \jr{J.\ of Supercond.} \textbf{10}, 393 (1997).


\bibitem{Taniguchi2015}
 \textsc{H.~Taniguchi},  \textsc{K.~Nishimura},  \textsc{S.~Goh},
  \textsc{S.~Yonezawa},  and  \textsc{Y.~Maeno},
 \jr{J.\ Phys.\ Soc.\ Jpn.} \textbf{84}, 014707 (2015).


\bibitem{Li2016}
 \textsc{J.~Li},  \textsc{D.~Zhao},  \textsc{Y.~Wu},  \textsc{S.~Li},
  \textsc{D.~Song},  \textsc{L.~Zheng},  \textsc{N.~Wang},  \textsc{X.~Luo},
  \textsc{Z.~Sun},  \textsc{T.~Wu},  and  \textsc{X.~Chen},
 \jr{arXiv:}, 1611.04694v1 (2016).


\bibitem{FukazawaPreprint}
 \textsc{H.~Fukazawa},  \textsc{Y.~Yamada},  \textsc{K.~Kondo},
  \textsc{T.~Saito},  \textsc{Y.~Kohori},  \textsc{K.~Kuga},
  \textsc{Y.~Matsumoto},  \textsc{S.~Nakatsuji},  \textsc{H.~Kito},
  \textsc{P.~Shirage},  \textsc{K.~Kihou},  \textsc{N.~Takeshita},
  \textsc{C.\,H. Lee},  \textsc{A.~Iyo},  and  \textsc{H.~Eisaki},
 \jr{J.\ Phys.\ Soc.\ Jpn.} \textbf{78}, 083712 (2009).


\bibitem{Korshunov2016}
 \textsc{M.\,M. Korshunov},  \textsc{Y.~Togushova},  and
  \textsc{O.~Dolgov},
 \jr{Phys.\ Usp.} \textbf{186}, 1315 (2016).


\end{thebibliography}

\end{document}